\documentclass[prd,showpacs,preprintnumbers,amsmath,amssymb,twocolumn,superscriptaddress,notitlepage]{revtex4-1}
\usepackage{graphicx,color}
\usepackage{graphicx}
\usepackage{amsmath,amssymb,amsfonts}
\usepackage{stackrel}
\usepackage{enumitem}
\usepackage[english]{babel}
\usepackage{verbatim}
\usepackage{comment}
\usepackage{ulem}
\usepackage{xcolor}
\usepackage{subcaption}
\usepackage{dcolumn}
\usepackage{physics}
\usepackage{verbatim}
\usepackage{float}
\usepackage{color}
\usepackage{multirow}
\usepackage{graphicx}
\usepackage{epstopdf}
\usepackage{epsfig}
\usepackage{slashed}
\usepackage{physics}
\usepackage{multirow}
\usepackage{makecell}
\usepackage{placeins}
\usepackage{setspace}
\usepackage{tabularx}
\usepackage{colortbl}
\usepackage{float}
\usepackage{placeins}
\usepackage{ragged2e}

\definecolor{darkred}{rgb}{.7,.1,.1}
\usepackage{hyperref}
\hypersetup{colorlinks=true,
	breaklinks=true,
	pdfstartview=Fit,
	linkcolor=blue,
	citecolor=blue,
	urlcolor=blue}
\usepackage{cleveref}

\graphicspath{
  {./}
  {figures/}
}

\renewcommand{\justify}{\leftskip=0pt \rightskip=0pt plus 0cm}

\definecolor{dark-green}{rgb}{0.1,0.7,0.3}

\begin{document}

\title{Implications of the KM3NeT Ultrahigh-Energy Event on Neutrino Self-Interactions}

\author{Yuxuan He}
\email{he.yx25@cityu.edu.hk}
\affiliation{Department of Physics, City University of Hong Kong, Kowloon, Hong Kong SAR, China}
\author{Jia Liu}
\email{jialiu@pku.edu.cn}
\affiliation{School of Physics and State Key Laboratory of Nuclear Physics and Technology, Peking University, Beijing 100871, China.}
\affiliation{Center for High Energy Physics, Peking University, Beijing 100871, China}
\author{Xiao-Ping Wang}
\email{hcwangxiaoping@buaa.edu.cn}
\affiliation{School of Physics, Beihang University, Beijing 100083, China}
\author{Yi-Ming Zhong}
\email{yiming.zhong@cityu.edu.hk}
\affiliation{Department of Physics, City University of Hong Kong, Kowloon, Hong Kong SAR, China}

\date{\today}
\preprint{$\begin{gathered}\includegraphics[width=0.05\textwidth]{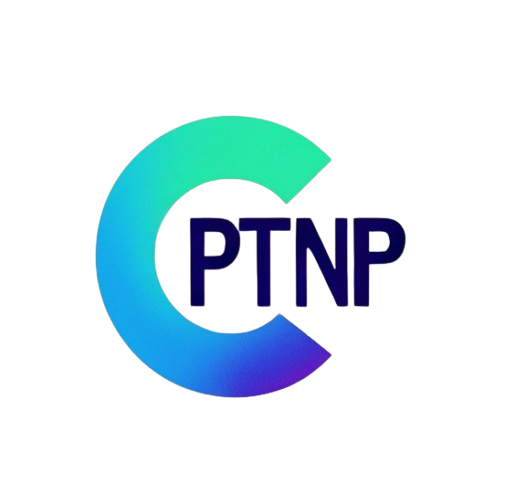}\end{gathered}$\, CPTNP-2025-007}

\begin{abstract}
Neutrino self-interactions ($\nu$SI) mediated by light bosonic particles can produce characteristic spectral dips in astrophysical neutrino fluxes, thereby altering the expected energy spectrum. The high-energy astrophysical neutrino spectrum has been extensively used to probe $\nu$SI models through these distinctive features. The recent detection of the ultrahigh-energy event KM3-230213A presents a new opportunity to explore $\nu$SI phenomenology at extreme energies. In this work, we investigate two implications of this observation, assuming the event originates from a diffuse power-law spectrum. First, we find that $\nu$SI-induced spectral distortions can mildly alleviate the tension between the KM3-230213A detection and the previous non-observation of PeV-scale neutrinos in IceCube data. Second, we derive the strongest constraints on the $\tau$-flavored $\nu$SI coupling strength for mediator masses around 100 MeV. Our analysis shows that neutrino telescopes can surpass existing collider bounds in this mass range. In the near future, IceCube-Gen2 is expected to significantly enhance $\nu$SI sensitivity, including regions relevant to alleviating the Hubble and neutrino mass tensions.
\end{abstract}

\pacs{}
\maketitle

\noindent 
\section{Introduction}
The neutrino sector of the Standard Model (SM) remains largely unexplored, offering significant opportunities to probe physics beyond the Standard Model (BSM). $\nu$SI have gained attention due to their potential to advance both particle physics and cosmology~\cite{Berryman:2022hds}. These interactions naturally arise in neutrino mass models, particularly those involving Majorons~\cite{Gelmini:1980re}, and in several gauge extensions of the SM~\cite{Bauer:2018onh}, making them a key focus of theoretical and experimental research.
$\nu$SI can affect cosmological processes by altering the free-streaming length of neutrinos, which influences the Cosmic Microwave Background (CMB) and matter power spectrum~\cite{Das:2020xke, Brinckmann:2020bcn}. Notably, these modifications could alleviate the Hubble ($H_0$) tension between high- and low-redshift measurements~\cite{DiValentino:2025sru} and address discrepancies in neutrino mass estimates from DESI~\cite{DESI:2025ejh} and direct experiments~\cite{Poudou:2025qcx}. Astrophysical signals can also probe $\nu$SI, such as their impact on core-collapse supernovae~\cite{Fuller:1988ega, Kachelriess:2000qc, Farzan:2002wx, Das:2017iuj, Chang:2022aas, Heurtier:2016otg, Fiorillo:2022cdq, Telalovic:2024cot, Akita:2022etk, Akita:2023iwq,Fiorillo:2023cas}.
Astrophysical neutrinos scattering off the cosmic neutrino background (C$\nu$B) can produce dips in the neutrino spectrum, revealing resonance at the mediator mass~\cite{Creque-Sarbinowski:2020qhz, Shoemaker:2015qul, Fiorillo:2020jvy, Fiorillo:2020zzj, Ng:2014pca, Farzan:2014gza, Bustamante:2020mep, Wang:2025qap, Mazumdar:2020ibx, Leal:2025eou, Borah:2025igh,Doring:2023vmk}. These spectral features can serve as a probe for $\nu$SI. Additionally, $\nu$SI mediator masses can be constrained by Big-Bang Nucleosynthesis (BBN), which affects the effective number of neutrino species ($N_{\text{eff}}$) \cite{Das:2020xke,Das:2022xsz, Brinckmann:2020bcn, Li:2023puz,Huang:2021dba,Ioka:2014kca}, and tested in terrestrial experiments by producing mediator particles or exploring new decay channels for mesons or $Z$ bosons \cite{Deppisch:2020sqh, Brdar:2020nbj, Dev:2024twk}.

Recently, the Cubic Kilometre Neutrino Telescope (KM3NeT) reported an energetic muon track, likely from a neutrino with energy $E_\nu = 220^{+570}_{-110} \ \text{PeV}$, known as KM3-230213A~\cite{KM3NeT:2025npi}. This has sparked studies on potential sources, including decaying dark matter~\cite{Jho:2025gaf, 
 Jiang:2025blz, Kohri:2025bsn, Khan:2025gxs, Murase:2025uwv, Barman:2025hoz, Borah:2025igh}, neutron portals~\cite{Alves:2025xul}, dark radiation~\cite{Narita:2025udw}, primordial black holes~\cite{Klipfel:2025jql, Jiang:2025blz,Choi:2025hqt}, and others~\cite{KM3NeT:2025bxl, Dzhatdoev:2025sdi, Neronov:2025jfj, Zhang:2025abk, Brdar:2025azm}. Some works also investigate associated gamma-ray emission~\cite{Crnogorcevic:2025vou,Fang:2025nzg,KM3NeT:2025aps,Das:2025vqd} and potential Lorentz violations~\cite{Yang:2025kfr,Cattaneo:2025uxk,KM3NeT:2025mfl}. However, IceCube~\cite{IceCube:2024fxo} and Pierre Auger~\cite{PierreAuger:2023pjg}, despite larger exposures, have not detected such high-energy neutrinos, only setting upper limits, which creates a tension with the KM3NeT event. If the KM3NeT event comes from a diffuse power-law source like those measured by IceCube~\cite{Abbasi:2021qfz,IceCube:2024fxo,IceCube:2020wum,IceCube:2025ary}, the tension could be as large as 3$\sigma$ \cite{Li:2025tqf, KM3NeT:2025ccp}, though assuming transient sources may reduce it to 2$\sigma$ \cite{Li:2025tqf}.

In this work, we explore the potential implications of the ultrahigh-energy KM3NeT event on $\nu$SI. First, $\nu$SI can induce dips in the neutrino flux spectrum, It can alter the high-energy neutrino (HE, referred as $10\,\text{TeV}< E_\nu < 10\,\text{PeV}$) and ultrahigh-energy neutrino (UHE, referred as $E_\nu > 10\,\text{PeV}$) flux and then potentially alleviate the tension  between the KM3NeT
event and IceCube measurements. Second, the UHE event opens new possibilities for heavier mediators, which could produce dips in the KM3NeT energy range—possibly constrained if the event is from a power-law source. Future observations, such as IceCube-Gen2~\cite{IceCube-Gen2:2020qha, Meier:2024flg}, could offer stronger, more reliable constraints. 

\section{$\nu$SI model}
In this work, we focus on $\nu$SIs mediated by a scalar $\phi$. The Lagrangian for this scenario can be written as 
\begin{equation}
\mathcal{L}_{\nu\text{SI}} =  \frac{1}{2} (\partial \phi)^2-\frac{1}{2}m_\phi \phi^2 -\frac{1}{2}\sum_{\alpha \beta} g_{\alpha, \beta} \bar \nu_\alpha \nu_\beta  \phi ,
\label{eq:lag}    
\end{equation}
where the indices $\alpha, \beta$ run over flavors ${e, \ \mu, \ \tau}$,  and $m_\phi$ is the mass of mediator $\phi$. $\phi$ interacts with the neutrino fields $\nu_\alpha$ and $\nu_\beta$ with the coupling strength $g_{\alpha \beta}$. The factor of $1/2$ before the coupling reflects the normalization choice to match the neutrino Majorana masses. When transforming to the neutrino mass basis, the coupling $g_{i,j}$ can be expressed as $g_{i,j} = \sum_{\alpha, \beta } U^*_{\alpha i} U_{\beta ,j} g_{\alpha, \beta}$, where $U_{\alpha i}$ represents the elements of the neutrino mixing matrix.

The Lagrangian Eq.\eqref{eq:lag} is not gauge-invariant and should be viewed as an effective low-energy description of a ultraviolet (UV) complete model. While several UV models incorporating $\nu$SI have been proposed~\cite{Blinov:2019gcj, Blum:2014ewa, Berryman:2018ogk, Kelly:2020pcy}, we do not discuss their constraints here, as our focus is not on any specific model.

Experimental constraints on $\nu$SI have been explored. For example, scalar mediators coupling to electron and muon neutrinos can increase meson invisible decay widths. For mediator masses below 100 MeV, meson decay limits set couplings at $g_{\mu\mu} \lesssim 10^{-2}$ and $g_{ee} \lesssim 10^{-3}$. The strongest terrestrial constraint on the $\tau\tau$ coupling comes from $Z$ boson invisible decays, with a limit of $g_{\tau\tau} \lesssim 0.2$\cite{Deppisch:2020sqh, Brdar:2020nbj, Dev:2024twk}. For mediators lighter than a few MeV, BBN constrains both $g_{\alpha\beta}$ couplings, as the mediator $\phi$ could affect $N_{\text{eff}}$ in the early universe\cite{Das:2020xke, Brinckmann:2020bcn}. Based on these limits, and we are investigating astrophysical neutrinos, we focus on the scenario where only $g_{\tau\tau}$ is non-zero and assume neutrinos reach Earth in their mass eigenstates due to oscillations. 

 \section{$\nu$SI effects on neutrino spectra}
$\nu$SI can significantly affect astrophysical neutrino observations by introducing features in the energy spectrum. When neutrinos scatter off the C$\nu$B, their energy can change during propagation.

If the center-of-mass energy of such scattering processes is near the resonant mass of the mediator, the s-channel scattering cross-section can be significantly enhanced. This enhancement leads to a reduction in the neutrino flux at certain energies, resulting in dips in the spectrum at those energies. The positions of these dips are determined by the mediator mass $m_\phi$ and the neutrino mass, according to the relation
$E_{\text{dip}} = m_\phi^2/\left(2m_\nu\right)$. 
The depth of these dips is primarily controlled by the width of the mediator and its couplings, with the width of the dips being dominated by the mediator's decay width. This qualitative description can be further quantified by solving the transport equation for the evolution of the neutrino energy spectrum~\cite{Esteban:2021tub}:
\begin{equation}
  \begin{aligned}
   \frac{ \partial  n_i (t, E_\nu)}{\partial t} &= \frac{\partial }{\partial E_\nu} [H(t) E_\nu n_i(t, E_\nu)]+\mathcal L_i (t ,E_\nu)  \\&-n_i(t, E_\nu) \sum_{j} N_j^b \sigma_{ij}(E_\nu) \\&
   +\sum_{jkl} N_j^b \int_{E_\nu}^\infty dE_\nu^\prime n_k(t, E_\nu^\prime) \frac{\sigma_{jk\to il}}{dE_\nu}(E_\nu,E_\nu^\prime),
\end{aligned} 
\label{eq:trans}
\end{equation}
where $n_i(t, E_\nu)=d N_i/d E_\nu$ represents the neutrino number density per energy in mass eigenstates. The first term on the right-hand side accounts for the propagation of astrophysical neutrinos, including redshift effects due to cosmic expansion. The second term encodes the production of astrophysical neutrinos, modeled here as being proportional to the star formation rate at the production redshift, with a power-law energy spectrum of the form $E_{\nu}^{-\gamma}$, with spectrum index $\gamma$ as a function of energy. The flavor compositions of the neutrino sources are assumed to be $(n_{\nu_e}:n_{\nu_\mu}:n_{\nu_\tau}) =(1:1:1)$ after oscillation. It can be shown that small perturbations of the flavor composition do not change the final result much.  The third and fourth terms represent the absorption and regeneration of neutrinos as they propagate through the C$\nu$B, with $N_i^b \approx 8.7 \times 10^{-13}(1+z)^3  \text{eV}^3$ being the number density of C$\nu$B neutrinos.

The dominant contribution to the absorption and regeneration cross-section comes from the s-channel scattering between astro-neutrino and background neutrino around the resonant energy, which can be parameterized as~\cite{Esteban:2021tub} 
\begin{equation}
    \sigma_{ij} ^s = \frac{|g_{ij}|^4}{16\pi} \frac{s}{(s-m_\phi^2)^2 +m_\phi^2 \Gamma_\phi^2},
\end{equation}
where $s$ is the center-of-mass energy, and $\Gamma_\phi$ is the decay width of the mediator $\phi$. When $s$ approaches $m_\phi^2$, the cross-section peaks at approximately $\sigma_{ij}\approx 1/( m_\phi^2)$, with width $\Gamma_\phi \approx (g^2 /16\pi) m_\phi $. This resonant energy corresponds to the astrophysical neutrino energy
$E_\nu =m_\phi^2 /\left(2 m_i\right)$, 
where $m_i$ is the i-th neutrino mass. Other scattering channels also contribute to the evolution of the neutrino flux during propagation, particularly when the coupling constant $g$ is large (i.e., $g \gg 10^{-2}$).

To quantitatively evaluate the neutrino flux for a given $\nu$SI scenario, we solve Eq.\eqref{eq:trans} with the initial condition $n_i(z \to \infty, E_\nu) = 0$, assuming no astrophysical sources in the early universe. This allows us to obtain the current neutrino flux, $n_i(z = 0, E_\nu)$. The evolution of the flux can be computed using the numerical code \texttt{nuSIprop}\cite{Esteban:2021tub}.

\section{$\nu$SI fits}.
To study the impact of $\nu$SI on the high- and ultrahigh-energy neutrino spectrum, we perform a joint fit using both HE and UHE neutrino data. In this energy range, KM3NeT reported a single event, KM3-230213A, with an energy of $E_\nu = 220^{+570}_{-110} \text{PeV}$ \cite{KM3NeT:2025npi}. Meanwhile, the IceCube Extremely-High-Energy (IceCube-EHE) sample and Auger reported no detections of neutrinos at these high energies \cite{PierreAuger:2023pjg, IceCube:2018fhm}. The fit for the KM3-230213A event with a single power law spectrum reveals a 3$\sigma$ tension between these UHE neutrino observations. In contrast, the $\nu$SI scenario offers a spectrum with potential dips within the observed energy range. 

We extend our analysis by including IceCube’s Northern-Sky Tracks (NST) \cite{Abbasi:2021qfz} and Enhanced Starting Track Event Selection (ESTES) \cite{IceCube:2024fxo} data, covering an energy range from TeV to EeV. We use two types of data: first, segmented power-law fits to the neutrino energy flux, where each energy interval is fitted with an $E^{-2}$ power-law, yielding a fiducial flux $\phi_{0,i}$ and associated errors; and second, likelihood functions from single power-law fits to each dataset.

For the joint fit, we make the simple assumption that the high-energy and ultrahigh-energy neutrinos originate from diffuse sources with fluxes following a single power-law spectrum:
$\Phi=\phi_0\left(E_\nu / E_0\right)^{-\gamma}$, 
where $\phi_0$ is the fiducial flux, $\gamma$ is the spectral index, and $\nu$SI is mediated by a scalar with mass $m_\phi$ and coupling to tau neutrinos $g_{\tau\tau}$. We further assume the neutrino mass sum to be $\sum  m_\nu =0.1 \ \text{eV}$\footnote{Recent DESI DR2~\cite{DESI:2025ejh} reported a constraint on sum of neutrino mass to be $\sum  m_\nu < 0.0642 \ \text{eV}$ when assuming $\Lambda$CDM model. But its best fit is when allowing dark energy to vary. Under the latter scenario, the neutrino total mass constraints relaxed to $\sum  m_\nu < 0.163 \ \text{eV}$. This allows us to use 0.1 eV as our benchmark value.} and assume the mass follows the normal hierarchy. The effect of $\nu$SI is computed using the numerical code \texttt{nuSIprop} \cite{Esteban:2021tub}, with neutrino mixing parameters from Ref. \cite{Esteban:2020cvm}.
 
We employ two approaches for constructing the likelihood functions for IceCube’s high-energy datasets. First, we use the segmented fluxes from the NST and ESTES datasets, listed as ``seg." in Table \ref{tab:fitres}. Second, we apply the likelihood functions based on single power-law fits, listed as ``single" in Table \ref{tab:fitres}, which is described in Refs.~\cite{Abbasi:2021qfz, IceCube:2024fxo}. The latter approach is relevant only if the energies of dips induced by $\nu$SI are higher than the fit range of the single power-law model.

The posterior distribution is computed by sampling the likelihood over the prior distributions of the model parameters. The chosen priors are as follows: The flux $\phi_0$ is assumed to follow a uniform distribution in logarithmic scale, $\mathcal{U}_{[-22,-15]}(\log_{10}(\phi_0 \ \text{GeV cm}^2 \text{s sr}))$; The spectral index $\gamma$ is assigned a uniform distribution, $\mathcal{U}_{[1,4]}(\gamma)$. The mediator mass $m_\phi$ follows a uniform distribution in logarithmic scale, $\mathcal{U}_{[5,9]}(\log_{10}(m_\phi/\text{eV}))$. The coupling strength $g_{\tau\tau}$ is modeled by a uniform distribution in logarithmic scale, $\mathcal{U}_{[-4,0]}(\log_{10}g_{\tau\tau})$.

\begin{table}[!htb]
\tabcolsep=4pt
\begin{center}
\resizebox{\columnwidth}{!}{%
\renewcommand{\arraystretch}{1.6}
\begin{tabular}{ c|c c c c}
\hline
\hline
 Data sets (UHE) & $-\log_{10}\left[\frac{\phi_0}{\text{GeV}^{-1} \text{cm}^{-2} \text{s}^{-1}\text{sr}^{-1}}\right]$ & $\gamma$ & $\log_{10}\left[\frac{m_\phi}{\text{MeV}}\right] $ & $\log_{10}g_{\tau\tau}$\\ \hline \hline
 + N seg. & $17.82^{+0.96}_{-0.01}$ &$2.32^{+0.39}_{-0.12}$ & $1.30^{+1.08}_{-1.39}$ & $0.68^{+2.81}_{-0.00}$\\ \hline
  + E seg. & $17.77^{+0.10}_{-0.03}$ & $2.54^{+0.18}_{-0.02}$& $1.27^{+1.26}_{-1.21}$ &$0.85^{+2.63}_{-0.00}$ \\ \hline
  + N single &$17.85^{+0.06}_{-0.06}$ & $2.33^{+0.11}_{-0.01}$& $1.23^{+1.00}_{-0.09}$ &$0.65^{+1.12}_{-0.19}$ \\ \hline
  + E single &$17.76^{+0.07}_{-0.02}$ & $2.56^{+0.08}_{-0.05}$& $1.22^{+1.37}_{-0.00}$ &$0.79^{+2.50}_{-0.04}$ \\ \hline
\hline
\end{tabular}
}
\end{center}
\caption{Best-fit values and uncertainties of parameters from various datasets: +N (UHE+NST), +E (UHE + ESTES) for combined fits of high- and ultrahigh-energy neutrino fluxes; “seg.” for segmented IceCube-HE flux; “single” for single power-law likelihood. 
}
\label{tab:fitres}
\end{table}

To sample the posterior, we use the \texttt{UltraNest} \cite{Buchner:2021cql} and \texttt{GetDist} \cite{Lewis:2019xzd} to estimate the posterior and compute the best-fit values of the parameters. The best-fit results for each parameter are presented in Table \ref{tab:fitres}, where each column corresponds to a different parameter, and each row corresponds to a dataset with specified approach.

\begin{figure}
    \centering
    \includegraphics[width=0.99\linewidth]{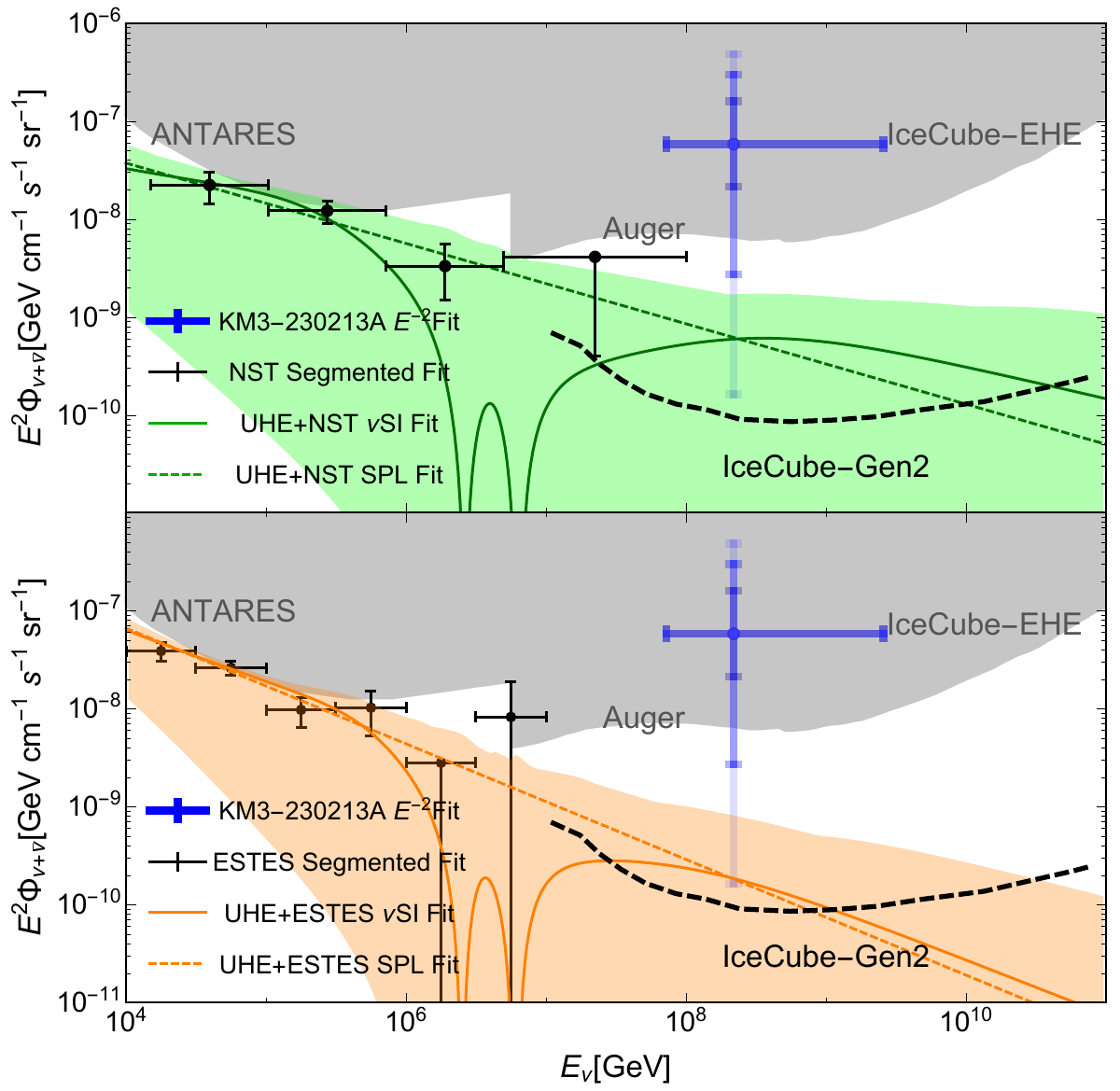}
    \caption{\justify In both panels, solid colored lines show the fitted flux with $\nu$SI effects, while dashed lines represent the fits without $\nu$SI. The colored bands indicate $1\sigma$ uncertainties from joint fits with $\nu$SI. Black crosses represent the segmented flux from observations, and the blue cross denotes the KM3NeT event. The gray shaded region shows flux limits from various observations~\cite{ANTARES:2024ihw, PierreAuger:2023pjg, IceCube:2018fhm}, and the black dashed lines indicate projected IceCube-Gen2 reach \cite{IceCube-Gen2:2020qha, Meier:2024flg}.
    }
    \label{fig:flux}
\end{figure}

To further evaluate the $\nu$SI scenario, we perform a posterior predictive check (PPC) to quantify the tension between the IceCube-EHE, Auger constraints, and the KM3-230213A event. The results are then converted to $z$-scores, as shown in Table \ref{tab:PPC}. These results indicate that the addition of $\nu$SI only slightly reduces the tension between the data, from a 3$\sigma$ tension to a 2.6–2.8$\sigma$ range. This relaxation is likely due to the fact that $\nu$SI can soften the neutrino spectrum, which increases the flux below the resonance. This effect allows for a smaller spectral index, thereby enhancing the flux at UHE and helping to alleviate the tension. The difference in the spectral indices can be observed in Fig. \ref{fig:flux} by comparing the slopes of the spectra fitted with and without $\nu$SI effects. 

However, since the above effect primarily alleviates the tension between the observed high-energy neutrino spectra in IceCube and the KM3NeT event, the tension due to the non-observation of ultra-high-energy neutrinos in IceCube and Auger at such energies remains unresolved. The tension between the prior non-observation of ultra-high-energy neutrinos and the KM3NeT event contributes approximately 2.5$\sigma$, and this cannot be resolved by adding $\nu$SI. Further illustrations of the event numeber and posterior can be found in Appendix \ref{post}.

In Bayesian statistics, model comparison is often quantified using Bayes factors, which provide a measure of the relative preference for one model over another. The Bayes factor is computed as the ratio of the evidence for two models, specifically
$\mathcal B =\mathbb E_{\nu \text{SI}} /\mathbb E_{\text{SPL}}$, 
where the evidence $\mathbb{E}$ for each model is defined as the integral of the posterior distribution over the parameter space:
$\mathbb E = \int d\Theta P(\Theta)$. 
We compute the Bayes factors for the $\nu$SI model relative to the single power-law (SPL) model using different neutrino datasets, employing the \texttt{UltraNest} sampler~\cite{Buchner:2021cql}. The results are summarized in Table~\ref{tab:PPC}.
\begin{table}[!htb]
\centering
\renewcommand{\arraystretch}{1.0}
\begin{tabular}{ c|c c c c}
\hline
\hline
 Data sets (UHE)  & +N seg.  & +E seg. & +N single & +E single \\ \hline \hline
 PPC $z$-score&2.85$\sigma$& 2.65$\sigma$  & 2.87$\sigma$ & 2.80$\sigma$\\ \hline
 Bayes Factor & 0.99&0.82 & 3.16&1.36\\ \hline
\hline
\end{tabular}
\caption{\justify The PPC results and the Bayes factors of the fits for different datasets. The PPC results listed in the first row are converted into $z$-score, reflecting the tension of the model with respect to each dataset. The second row listed the Bayes factor for these fits. 
}
\label{tab:PPC}
\end{table}

We found that when fitting segmented data, there is no strong preference for the $\nu$SI model over the single power law. However, for the likelihood based on the single power-law fit for high-energy events, a slight preference for $\nu$SI is observed, particularly in the NST dataset.

The observed differences between datasets could arise from the inherent assumptions in the segmented fitting approach. Specifically, the segmented fitting assumes a constant $E^{-2}$ power law within each energy interval, which may not capture the true spectral behavior in a more realistic scenario.  When using the single power-law fit for events below the resonant energy, slight deviations persist between the single power-law and the $\nu$SI spectrum. These deviations suggest the need for more detailed analysis to fully account for the $\nu$SI effects, especially in the lower energy ranges.

In Fig. \ref{fig:flux}, we show the fluxes computed from the best-fit parameters for the joint fit of UHE observations and IceCube, NST, or ESTES datasets with ``single" approach. The solid lines represent the spectrum with the $\nu$SI effects included. The solid green line is for ``+ N single", while the solid orange line is for ``+ E single". The dashed lines show the fit for a single power-law spectrum without the $\nu$SI effect. The colored shaded regions correspond to the $1\sigma$ uncertainty bands for the $\nu$SI fits. The segmented flux used in the fit is indicated by black crosses. Additionally, we plot the projected reach of neutrino flux from IceCube-Gen2 \cite{IceCube-Gen2:2020qha, Meier:2024flg}, which provides an opportunity to test $\nu$SI effects in the future.

\section{Constraints on $\nu$SI parameters }
Despite exploring the possibility of alleviating the tension between different neutrino flux measurements and the KM3-230213A event through $\nu$SI, it is also worthwhile to use this highest-energy neutrino event to constrain the parameter space of the $\nu$SI scenario. Although only one event at such high energy has been observed so far, assuming that its flux originates from a diffuse neutrino source, meaningful constraints on $\nu$SI can still be obtained. If the energy of the spectral dip coincides with that of KM3-230213A, the expected event rate would be suppressed too severely to observe any events. The energy of KM3-230213A, approximately 100 PeV, allows us to constrain mediator masses around 100 MeV. The constraint is strongest at $m_\phi \approx \sqrt{2 m_\nu E_{\text{KM3}}}$. 

\begin{figure}[t]
    \centering
    \includegraphics[width=0.99\linewidth]{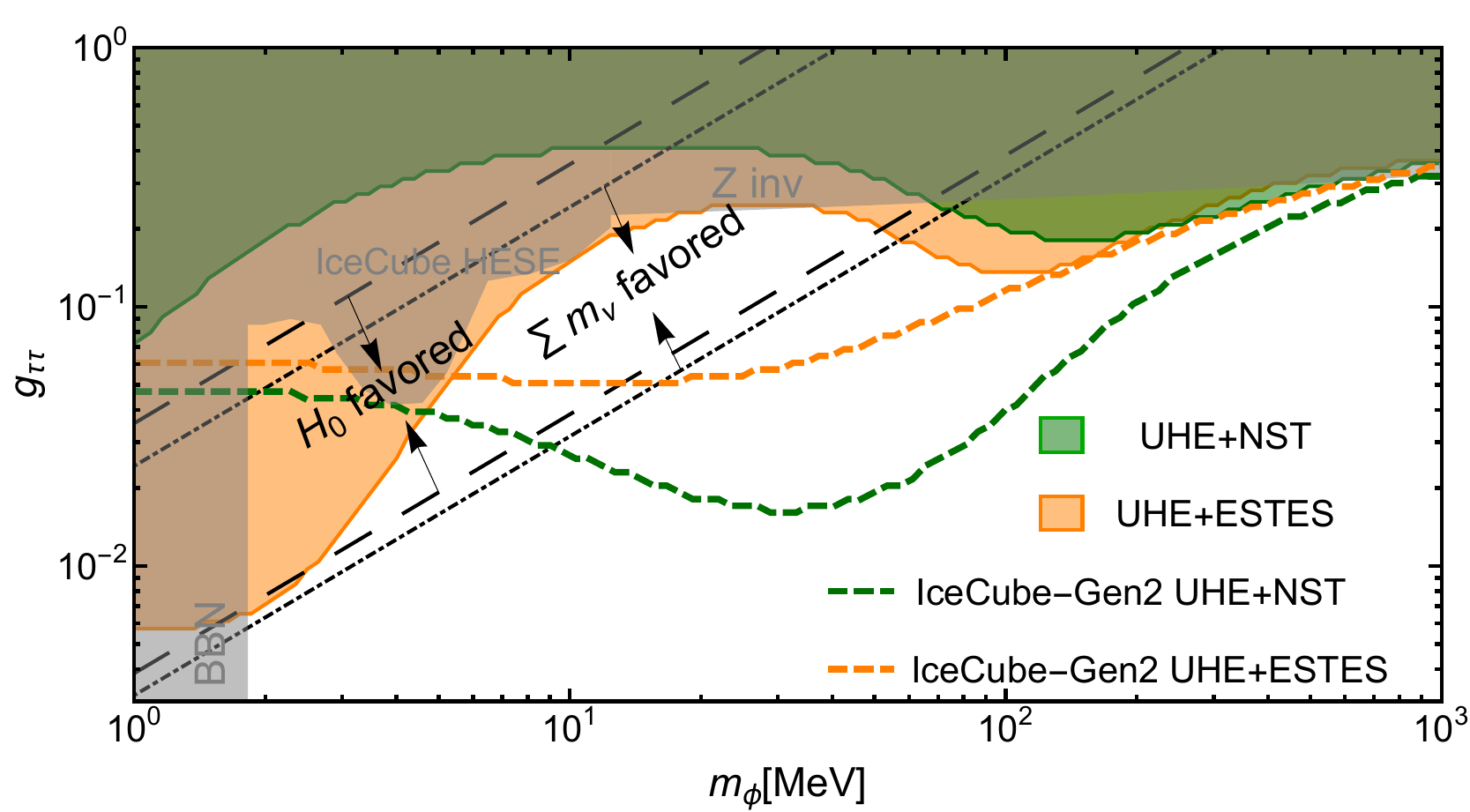}
    \caption{\justify The 95\% CL constraints on $\nu$SI mediator masses and couplings from the UHE+NST seg. (green shaded) and UHE+ESTES seg. (orange shaded) joint fits. The dashed lines show our projected exclusion for IceCube-Gen2 \cite{IceCube-Gen2:2020qha, Meier:2024flg} based on the UHE+NST and UHE+ESTES $\nu$SI-Fit flux parameters.The gray region marks exist exclusions from BBN, $Z$ boson decays, and IceCube HESE \cite{Brdar:2020nbj, Esteban:2021tub}. The region between the black dashed lines highlights the parameter space that may resolve the Hubble tension, and the dot-dashed lines indicate space that could address the neutrino mass tension \cite{Poudou:2025qcx}. 
    }
    \label{fig:cons}
\end{figure}

We obtain the 95\% confidence level constraint on the parameter $g_{\tau \tau}$ by marginalizing the previously defined posterior and computing the limit according to the following integral. Specifically, we determine $g_{\tau\tau}^{\text{95\%}}$ by solving the following equation:
\begin{equation}
    0.95 = \frac{\int_{0}^{g_{\tau\tau}^{\text{95\%}}} p(m_\phi, g_{\tau\tau}) dg_{\tau\tau}}{\int p(m_\phi, g_{\tau\tau}) dg_{\tau\tau}} ,
\end{equation}
where $p(m_\phi, g_{\tau\tau})$ denotes the posterior marginalized over other parameters. Using ultrahigh-energy data alone does not yield meaningful limits, as the expected flux can be very low. However, by combining it with the high-energy flux measured by IceCube and assuming that both originate from the same power-law source, we obtain a more stringent constraint. As shown in Fig.~\ref{fig:cons}, when combined with the ESTES dataset, the constraint around a mediator mass of 100 MeV becomes approximately twice as stringent as those from collider experiments. The strong constraints around the 100 MeV are mainly by virtue of the UHE event KM3-230213A. In Appendix \ref{compare}, we show the effect of this event by comparing the constraint with the result obtained by ignoring the KM3NeT event.

In Fig.~\ref{fig:cons}, we also show the constraints from $Z$ boson invisible decays, BBN~\cite{Brdar:2020nbj, Esteban:2021tub}, and previous IceCube HESE data~\cite{Esteban:2021tub} as the gray shaded region. Other constraints, such as those from $\tau$ decays~\cite{Brdar:2020nbj} or supernova observations~\cite{Chang:2022aas}, are either weaker than the $Z$ invisible decay limits or are model-dependent, and thus are not shown in the figure.

For the choice of neutrino masses, it is worth noting that we fix $\sum m_\nu = 0.1\ \text{eV}$ and use normal mass hierarchy best fit parameters in Ref.~\cite{Esteban:2020cvm} for our analysis, and the dips appear at $m_\phi = \sqrt{2 m_\nu E}$. Given that $\sum m_\nu$ has been strongly constrained by cosmological observations ($\sum m_\nu < 0.12 \ \text{eV}$) and neutrino oscillation experiments ($\sum m_\nu > 0.06 \ \text{eV}$), we expect that varying the neutrino mass within this narrow window induces only minor shifts in the dip locations and does not significantly affect our constraints.

These constraints rely on the hypothesis of a diffuse power-law flux shared between high-energy (IceCube) and ultrahigh-energy (KM3NeT) observations. The observed tension suggests that alternative origins (e.g., transient sources) could weaken these bounds, as explored in the Appendix \ref{tran}. Nonetheless, our results represent the strongest astrophysical probe in this mass range to date, assuming the power-law model.
Although our constraint depends on certain assumptions regarding the event’s origin, future ultrahigh-energy observations, such as those by IceCube-Gen2~\cite{IceCube-Gen2:2020qha, Meier:2024flg}, are expected to reveal the nature of ultrahigh-energy neutrino sources and provide more reliable probes of $\nu$SI. In Fig.~\ref{fig:cons}, we calculate the projected sensitivity of IceCube-Gen2, assuming the ultrahigh-energy neutrino originates from diffuse power-law sources. Dashed green and orange lines use $\nu$SI flux parameters shown in Fig.~\ref{fig:flux}. Details of calculations are provided in Appendix \ref{gen2}. In Fig.~\ref{fig:cons}, we show that UHE+ESTES outperforms UHE+NST for current IceCube data due to its larger high-energy flux, yielding a stronger bound with UHE events. In contrast, UHE+NST dominates in IceCube-Gen2 projections due to its higher flux above 3 PeV.

Regarding the potential solutions to the Hubble tension, two scenarios of $\nu$SI—the so-called ``strongly interacting neutrino mode" (SI$\nu$) and ``moderately interacting neutrino mode" (MI$\nu$)—were investigated~\cite{Kreisch:2019yzn}. However, CMB polarization and galaxy power spectrum measurements disfavor the SI$\nu$ scenario~\cite{Das:2020xke, Brinckmann:2020bcn, Camarena:2024daj,RoyChoudhury:2020dmd}. Therefore, in Fig.~\ref{fig:cons}, we show only the MI$\nu$ region, defined by $10^{-4.83} \leq g_{\tau\tau}^2 \ \text{MeV}^2/m_\phi^2 \leq 10^{-2.9}$ (bounded by the black dashed lines). The same figure also shows the MI$\nu$ parameter space that alleviates the \textit{neutrino mass tension} between DESI and oscillation experiments, characterized by $g_{\tau\tau}^2 \  \text{MeV}^2/m_\phi^2 \leq 10^{-3.23}$ (bounded by the black dash-dotted lines) \cite{Poudou:2025qcx}.
Note that Ref.~\cite{Poudou:2025qcx} does not specify a lower bound; therefore, we adopt the lower edge of its prior, $g_{\tau\tau}^2 \ \text{MeV}^2/m_\phi^2 \geq 10^{-5}$, as the lower limit shown. Overall, the parameter space that alleviates current cosmological tensions can be largely probed by future IceCube-Gen2 observations.

\section{Conclusion}
In this work, we discuss the possible implications of the UHE neutrino event KM3-230213A for $\nu$SIs. We focus on the scenario in which the UHE neutrino originates from a power-law source and subsequently scatters with the C$\nu$B during its propagation, leading to a dip in the energy spectrum of the neutrino flux, with only interactions involving tau neutrinos. First, we test whether the introduction of such spectral distortions can help alleviate the tension between IceCube high-energy neutrino observations and the KM3-230213A event. While the tension can be slightly reduced, but it remains above the $2.5\sigma$ level. Next, we compare the $\nu$SI scenario to a single power-law model by combining the UHE event with two IceCube high-energy datasets and calculating the Bayes factors. We find no strong preference for $\nu$SI, except when using a single power-law fit to KM3-230213A combined with the IceCube NST dataset, where a slight preference for $\nu$SI appears. Finally, we set constraints on the mediator couplings in the $\nu$SI scenario. Assuming the flux originates from a source fitted by the ESTES dataset, we constrain $g_{\tau\tau}$ at $m_\phi = 100\ \text{MeV}$ to the level of $10^{-2}$, approximately a factor of two stronger than the constraints from $Z$ boson invisible decays. 
In the near future, IceCube-Gen2 is expected to greatly extend $\nu$SI sensitivity, including regions that could alleviate the Hubble and neutrino mass tensions.

\begin{acknowledgements}
We thank Christina Gao, Siyang Ling, Kenny Ng, and Bei Zhou for useful discussions.  Y.H. and Y.Z. are supported by the GRF grants No. 11302824 and No. 11310925 from the Hong Kong Research Grants Council and the Grant No. 9610645 from the City University of Hong Kong. The work of J.L. is supported by the National Science Foundation of China under Grant No. 12235001, No. 12475103, and the State Key Laboratory of Nuclear Physics and Technology under Grant No. NPT2025ZX11.  The work of X.P.W. is supported by the National Science Foundation of China under Grant No. 12375095, and the Fundamental Research Funds for the Central Universities. J.L. and X.P.W. also thank APCTP, Pohang, Korea, for their hospitality during the focus program [APCTP-2025-F01], from which this work greatly benefited. The authors gratefully acknowledge the valuable discussions and insights provided by the members of the China Collaboration of Precision Testing and New Physics.
\end{acknowledgements}

\appendix 

\section{The construction of likelihoods}
\label{app:}
We construct the likelihood function as described in Ref.~\cite{KM3NeT:2025ccp}:
\begin{equation}
    \mathcal{L}(\phi_0,\gamma, m_\phi,g_{\tau\tau}) =\prod_{d}\mathcal{L}^d(\phi_0,\gamma, m_\phi,g_{\tau\tau}),
\end{equation}
where the superscript $d$ labels the datasets datasets (IC-EHE, Auger, KM3NeT, and one of the IceCube high-energy datasets). We avoid fitting the three IceCube high-energy datasets jointly due to non-trivial overlaps in energy, as stated in Ref.~\cite{KM3NeT:2025ccp}. Thus, for the UHE datasets ($d \in { \text{KM3NeT, Auger, IceCube-EHE} }$), the likelihood function is:
\begin{align}
    \mathcal{L}^d(\Theta) & = \mathcal{L}^d_{\text{in}}(\Theta)\mathcal{L}^d_{\text{out}}(\Theta), \\
       \mathcal{L}^d_\text{in/out}(\Theta) &= \text{Poisson} (N_{\text{obs}}^{d,\text{in/out}}, N_{\text{exp}}^{d,\text{in/out}}(\Theta)),
\end{align}
where $\mathcal{L}^d_{\text{in/out}}(\Theta)$ follows a Poisson distribution based on the observed and expected number of events inside  or outside the 2$\sigma$ interval for KM3-230213A:
\begin{equation}
    N_{\text{exp}}^{d,\text{in/out}}(\Theta) = \int_{\text{in/out}} dE \ \mathcal E^d(E) \Phi_{\bar \nu +\nu}(E,\Theta).
\end{equation}
For each observation, the effective area $ \mathcal E^d(E) $ is taken from Ref.~\cite{KM3NeT:2025ccp}, and we choose the in and out intervals following Ref.~\cite{KM3NeT:2025ccp}.

For the IceCube high-energy neutrino flux measurements, we utilize the segmented fluxes from the Northern-Sky Tracks (NST) and Enhanced Starting Track Event Selection (ESTES) datasets, along with likelihoods derived from single power-law fits as implemented in Refs.~\cite{Abbasi:2021qfz, IceCube:2024fxo}. For segmented fitting, we apply the following likelihood function \cite{Barlow:2004wg}
\begin{align}
 \log\mathcal{L} ={} \sum_{i} -\frac{1}{2} \left(\frac{\hat{x}_i - x_i}{\sigma_i + \sigma_i' (x_i - \hat{x}_i)}\right)^2,
\end{align}
where $x_i$ and $\hat{x}_i$ represent the observed and predicted fluxes integrated within each segment, respectively. The uncertainties $\sigma_i$ and $\sigma_i'$ are computed as:

\begin{equation} \sigma_i = \frac{2 \sigma_+ \sigma_-}{\sigma_+ + \sigma_-}, \quad \sigma_i' = \frac{\sigma_+ - \sigma_-}{\sigma_+ + \sigma_-}, 
\end{equation} 

where $\sigma_{\pm}$ denotes the upper and lower one-sigma deviations for the fluxes within each segment.

The PPC is calculated using the following $p$-value~\cite{KM3NeT:2025ccp}:

\begin{equation} 
p_{\text{PPC}} = \int P(\Theta, D) \prod_d \mathcal{L}^d_{\text{in}}(\Theta)  \text{d} \Theta, 
\end{equation} 
where $P(\Theta, D)$ represents the posterior of the parameters $\Theta$ given the datasets $D$. 

\section{Estimation of the IceCube-Gen2 projected sensitivity}
\label{gen2}
We generate two simulated datasets with power-law spectral shapes, adopting the flux normalization $\phi_0$ and spectral indices corresponding to the best-fit values of the UHE+NST and UHE+ESTES models from Table 1 of the main text.
More specifically, we first divide the energy range [3 PeV, 100 EeV] into ten logarithmic-even bins. The expected event number in each bin can be computed as
\begin{equation}
    N_i^{d} = \int_{E_i}^{E_{i+1}} dE \ \mathcal E(E) \Phi_{\bar \nu +\nu}^d(E),
\end{equation}
where $i$ runs from 1 to 10, stands for energy bins, and the effective area $\mathcal E(E)$ of IceCube-Gen2 can be estimated as 10 times that of IceCube. $d$ stands for which spectrum we use. Mock data in each bin follows a Poisson distribution with the expected event number given above.

Then we construct the likelihood function by taking the product of the Poisson distribution of each energy bin. Each Poisson distribution has the expected event number computed, including $\nu$SI effects, and the observed event numbers are the mock data. Constraints on $\nu$SI coupling are derived using the same profile likelihood method described in the main text.

\begin{figure*}
    \centering
    \includegraphics[width=0.48\linewidth]{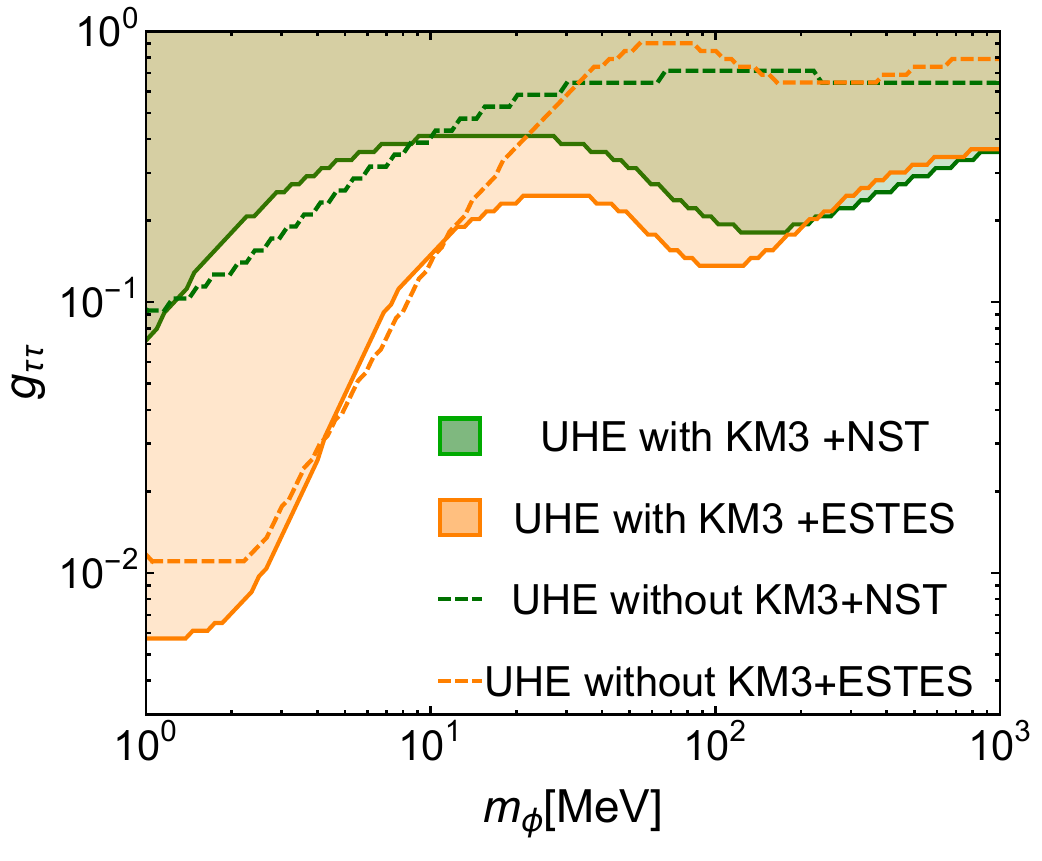}
    \includegraphics[width=0.48\linewidth]{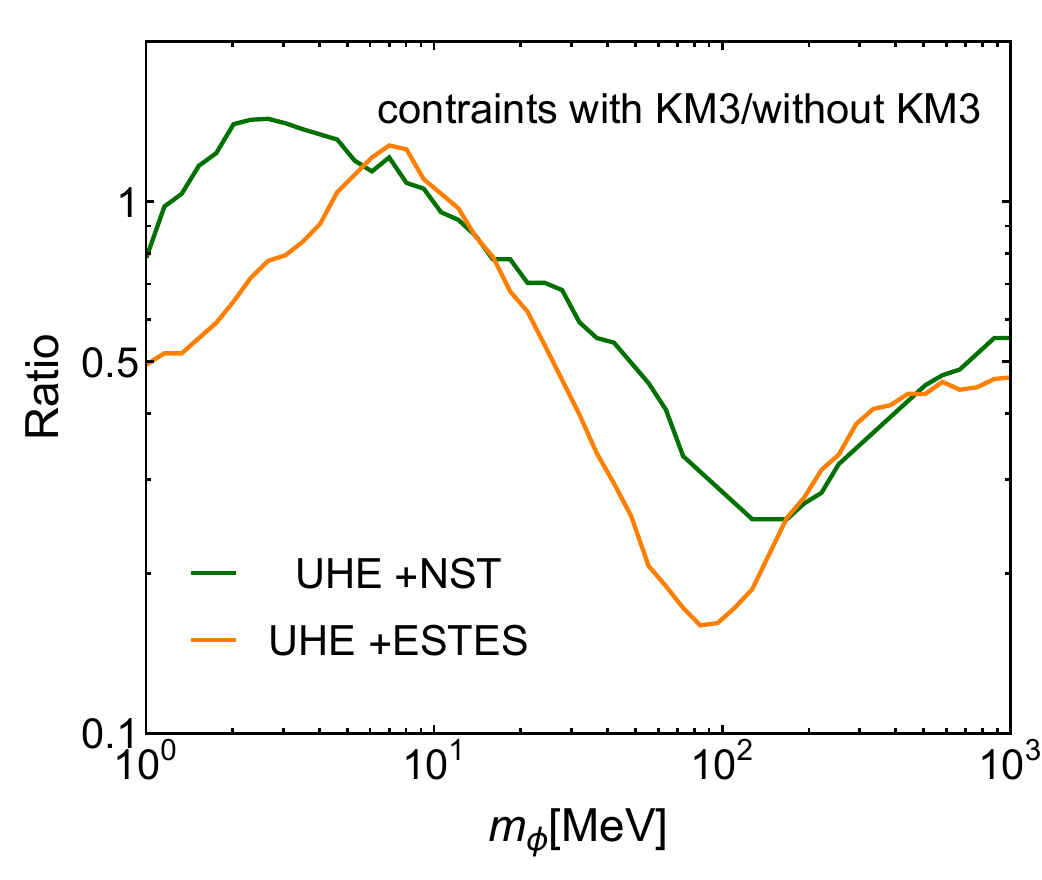}
    \caption{In the left panel of the plot, we show the constraints on $\nu$SI coupling derived without the KM3-230213A event, in dashed colored lines, while the original constraints are in shaded regions. In the right panel, the ratios of the constraints with and without the KM3-230213A event are shown in colored lines. }
    \label{fig:noKM}
\end{figure*}

\section{Illustration of the Posterior distributions}
\label{post}
\begin{figure}
    \centering
    \includegraphics[width=0.9\linewidth]{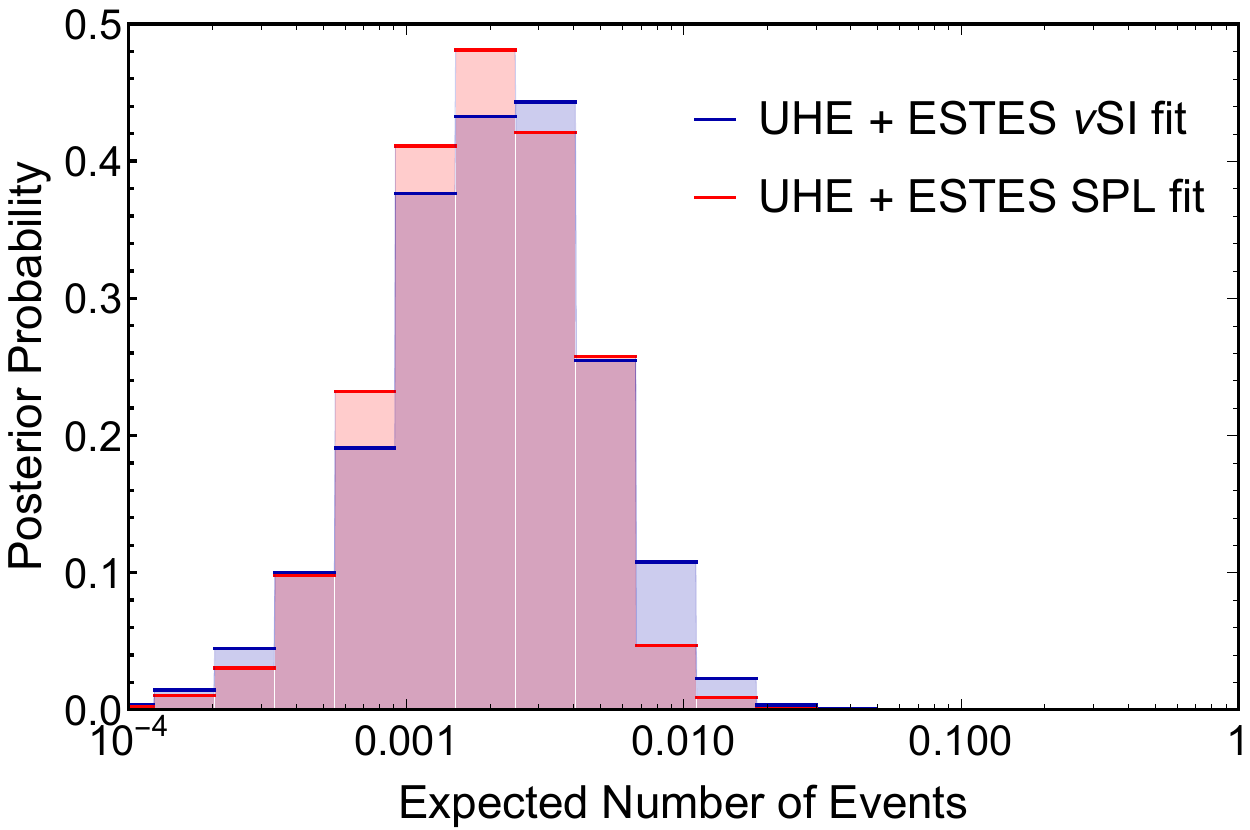}
    
    \caption{The posterior distribution of the KM3NeT event number expectation value with energy in the 1 $\sigma$ error bar range. The results for $\nu$SI and the SPL fits of the UHE + ESTES data set are shown in blue  red shaded, respectively. }
    \label{fig:relax1}
\end{figure}

\begin{figure}
    \centering
    \includegraphics[width=0.9\linewidth]{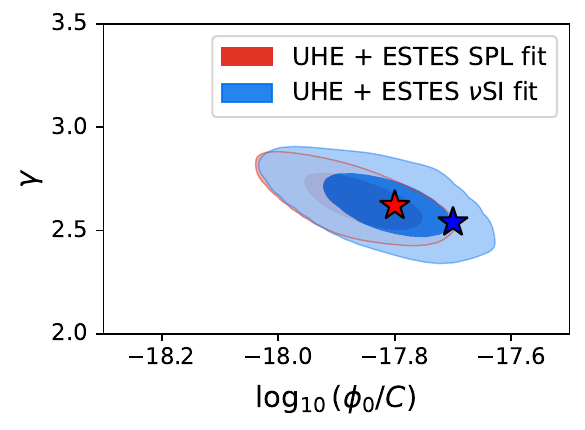}
    \caption{We show the $90$ \% and $95$ \% favored region in posterior of the flux intensities and the flux indices. The results for the $\nu$SI and the SPL fits of the UHE + ESTES data set are shown in blue shaded region and red shaded region. The stars mark the best fit points, respectively. }
    \label{fig:relax2}
\end{figure}

To illustrate the tension alleviation in $\nu$SI, In Fig. \ref{fig:relax1}, we plot the posterior distribution of the KM3NeT event number expectation value with energy in the 1 $\sigma$ error bar range. The $\nu$SI and SPL fits are illustrated in blue and red; the plot shows that the increase in the event number is modest. In Fig. 
\ref{fig:relax2}, the posterior of the flux intensities and the flux indices are plotted for the $\nu$SI and the SPL fits of the UHE + ESTES data set. The plot indicates that lower flux indices and higher intensities are permitted in $\nu$SI, which helps alleviate the tension between the IceCube results.

\section{Brief discussion of $\nu$SI effects for transient sources}
\label{tran}

For transient events, the tension can be reduced mainly due to the reduction of the integrated time. If the transient just begins to produce the neutrino flux at the same time when KM3NeT begins to collect data, the integrated times for KM3NeT and IceCube then turn out to be the same. Then the tension can only come from the difference of effective area, for this circumstance to be reduced to $2 \sigma$ \cite{Li:2025tqf}. If the transient lasts as long as the data collection time for IceCube, then the tension is still around 2.9$\sigma$ \cite{Li:2025tqf}, and not reduced very much. For our self-interaction neutrino scenario, the reduction of tension is mainly due to the reduction of the spectral index to produce more events at the tail while fitting the previously measured high-energy neutrino at the same time. This effect only matters when we assume the ultra-high-energy neutrino comes from the same diffuse source as the previously measured high-energy neutrinos. For the transient scenario, since we need not fit the total spectrum, the contribution to the expected event for IceCube and KM3NeT should be only proportional to their effective era and integrated time, hence not affected much by self-interacting neutrinos.

\begin{figure*}
    \centering
    \includegraphics[width=0.7\linewidth]{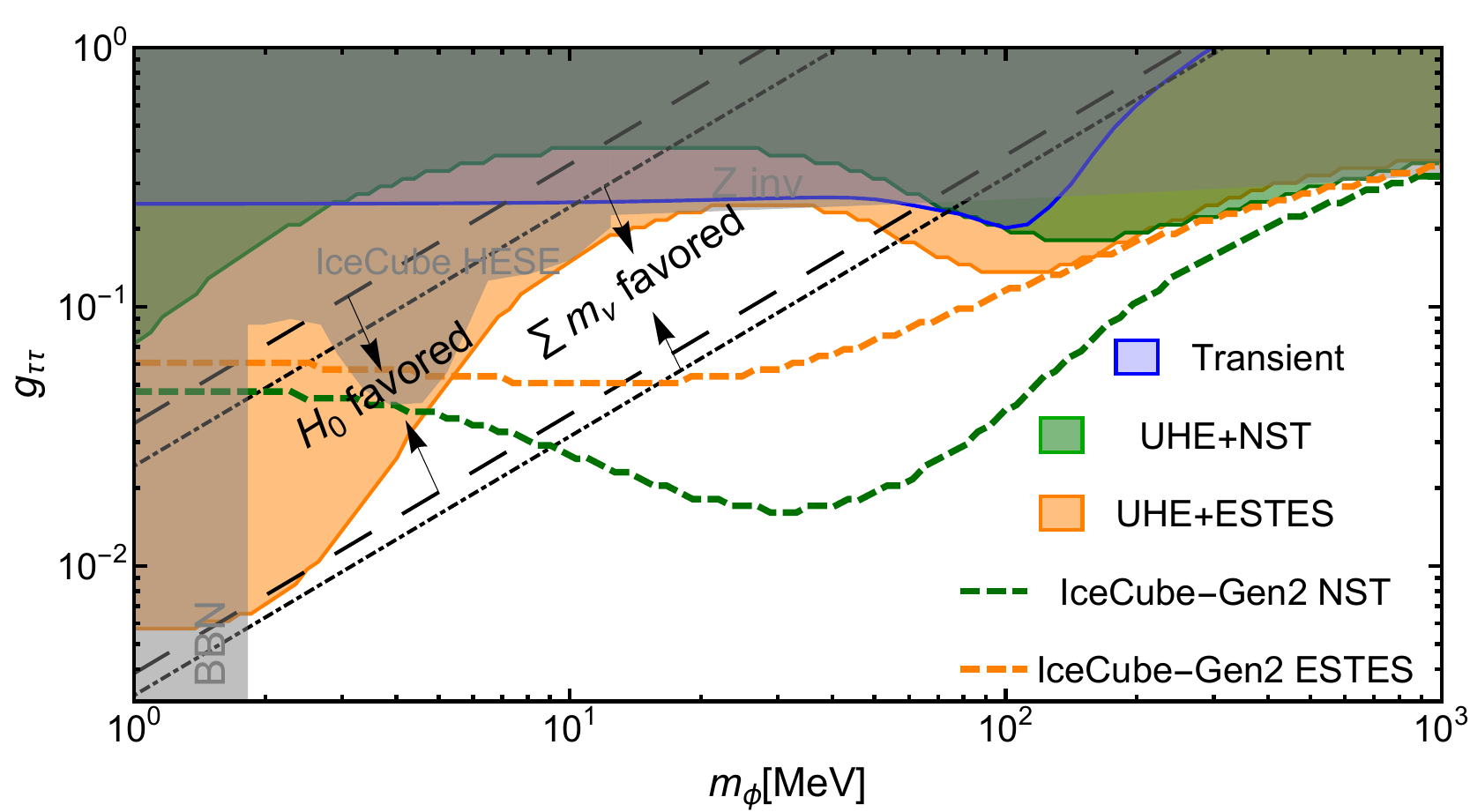}
    \caption{In the blue shaded region, we show the constraints obtained by assuming the KM3-230213A comes from a transient source with redshift $z=0.87$.}
    \label{fig:trans}
\end{figure*}

Although $\nu$SI will not reduce tension for the transient scenario, for transients with source distances not too close to us, i.e. $D > 0.1\ \text{Mpc}$, which is shorter than the mean free path, neutrinos from such transient sources can scatter with the cosmic neutrino background, and modify the flux. We can derive constraints when assuming a transient source with a large enough redshift. The constraint is obtained by solving the coupling $g_{\tau\tau}$ such that the flux is reduced to 10\% of the original one due to the $\nu$SI. We show the result in Fig. \ref{fig:trans}, where the constraint is shown in the blue shaded region.

\section{The Illustration of the impact of KM3-230213A on constraining $\nu$SI}
\label{compare}

In Fig. \ref{fig:noKM}, with colored dotted lines, we show the constraints obtained from HE and UHE datasets and remove the effect of the KM3NeT UHE event. We can see that, when compared with the shaded regions obtained by combining the KM3NeT UHE event and previous observations, the constraints in dashed lines are much weaker when the mediator mass is above $\mathcal O (10)$ MeV. It demonstrates that the KM3-230213A event brings us strong constraints for mediator mass above $\mathcal O (10)$ MeV.



\bibliography{reference}

\begin{thebibliography}{85}%
\makeatletter
\providecommand \@ifxundefined [1]{%
 \@ifx{#1\undefined}
}%
\providecommand \@ifnum [1]{%
 \ifnum #1\expandafter \@firstoftwo
 \else \expandafter \@secondoftwo
 \fi
}%
\providecommand \@ifx [1]{%
 \ifx #1\expandafter \@firstoftwo
 \else \expandafter \@secondoftwo
 \fi
}%
\providecommand \natexlab [1]{#1}%
\providecommand \enquote  [1]{``#1''}%
\providecommand \bibnamefont  [1]{#1}%
\providecommand \bibfnamefont [1]{#1}%
\providecommand \citenamefont [1]{#1}%
\providecommand \href@noop [0]{\@secondoftwo}%
\providecommand \href [0]{\begingroup \@sanitize@url \@href}%
\providecommand \@href[1]{\@@startlink{#1}\@@href}%
\providecommand \@@href[1]{\endgroup#1\@@endlink}%
\providecommand \@sanitize@url [0]{\catcode `\\12\catcode `\$12\catcode
  `\&12\catcode `\#12\catcode `\^12\catcode `\_12\catcode `\%12\relax}%
\providecommand \@@startlink[1]{}%
\providecommand \@@endlink[0]{}%
\providecommand \url  [0]{\begingroup\@sanitize@url \@url }%
\providecommand \@url [1]{\endgroup\@href {#1}{\urlprefix }}%
\providecommand \urlprefix  [0]{URL }%
\providecommand \Eprint [0]{\href }%
\providecommand \doibase [0]{https://doi.org/}%
\providecommand \selectlanguage [0]{\@gobble}%
\providecommand \bibinfo  [0]{\@secondoftwo}%
\providecommand \bibfield  [0]{\@secondoftwo}%
\providecommand \translation [1]{[#1]}%
\providecommand \BibitemOpen [0]{}%
\providecommand \bibitemStop [0]{}%
\providecommand \bibitemNoStop [0]{.\EOS\space}%
\providecommand \EOS [0]{\spacefactor3000\relax}%
\providecommand \BibitemShut  [1]{\csname bibitem#1\endcsname}%
\let\auto@bib@innerbib\@empty
\bibitem [{\citenamefont {Berryman}\ \emph {et~al.}(2023)\citenamefont
  {Berryman} \emph {et~al.}}]{Berryman:2022hds}%
  \BibitemOpen
  \bibfield  {author} {\bibinfo {author} {\bibfnamefont {J.~M.}\ \bibnamefont
  {Berryman}} \emph {et~al.},\ }\href
  {https://doi.org/10.1016/j.dark.2023.101267} {\bibfield  {journal} {\bibinfo
  {journal} {Phys. Dark Univ.}\ }\textbf {\bibinfo {volume} {42}},\ \bibinfo
  {pages} {101267} (\bibinfo {year} {2023})},\ \Eprint
  {https://arxiv.org/abs/2203.01955} {arXiv:2203.01955 [hep-ph]} \BibitemShut
  {NoStop}%
\bibitem [{\citenamefont {Gelmini}\ and\ \citenamefont
  {Roncadelli}(1981)}]{Gelmini:1980re}%
  \BibitemOpen
  \bibfield  {author} {\bibinfo {author} {\bibfnamefont {G.~B.}\ \bibnamefont
  {Gelmini}}\ and\ \bibinfo {author} {\bibfnamefont {M.}~\bibnamefont
  {Roncadelli}},\ }\href {https://doi.org/10.1016/0370-2693(81)90559-1}
  {\bibfield  {journal} {\bibinfo  {journal} {Phys. Lett. B}\ }\textbf
  {\bibinfo {volume} {99}},\ \bibinfo {pages} {411} (\bibinfo {year}
  {1981})}\BibitemShut {NoStop}%
\bibitem [{\citenamefont {Bauer}\ \emph {et~al.}(2018)\citenamefont {Bauer},
  \citenamefont {Foldenauer},\ and\ \citenamefont {Jaeckel}}]{Bauer:2018onh}%
  \BibitemOpen
  \bibfield  {author} {\bibinfo {author} {\bibfnamefont {M.}~\bibnamefont
  {Bauer}}, \bibinfo {author} {\bibfnamefont {P.}~\bibnamefont {Foldenauer}},\
  and\ \bibinfo {author} {\bibfnamefont {J.}~\bibnamefont {Jaeckel}},\ }\href
  {https://doi.org/10.1007/JHEP07(2018)094} {\bibfield  {journal} {\bibinfo
  {journal} {JHEP}\ }\textbf {\bibinfo {volume} {07}},\ \bibinfo {pages}
  {094}},\ \Eprint {https://arxiv.org/abs/1803.05466} {arXiv:1803.05466
  [hep-ph]} \BibitemShut {NoStop}%
\bibitem [{\citenamefont {Das}\ and\ \citenamefont
  {Ghosh}(2021)}]{Das:2020xke}%
  \BibitemOpen
  \bibfield  {author} {\bibinfo {author} {\bibfnamefont {A.}~\bibnamefont
  {Das}}\ and\ \bibinfo {author} {\bibfnamefont {S.}~\bibnamefont {Ghosh}},\
  }\href {https://doi.org/10.1088/1475-7516/2021/07/038} {\bibfield  {journal}
  {\bibinfo  {journal} {JCAP}\ }\textbf {\bibinfo {volume} {07}},\ \bibinfo
  {pages} {038}},\ \Eprint {https://arxiv.org/abs/2011.12315} {arXiv:2011.12315
  [astro-ph.CO]} \BibitemShut {NoStop}%
\bibitem [{\citenamefont {Brinckmann}\ \emph {et~al.}(2021)\citenamefont
  {Brinckmann}, \citenamefont {Chang},\ and\ \citenamefont
  {LoVerde}}]{Brinckmann:2020bcn}%
  \BibitemOpen
  \bibfield  {author} {\bibinfo {author} {\bibfnamefont {T.}~\bibnamefont
  {Brinckmann}}, \bibinfo {author} {\bibfnamefont {J.~H.}\ \bibnamefont
  {Chang}},\ and\ \bibinfo {author} {\bibfnamefont {M.}~\bibnamefont
  {LoVerde}},\ }\href {https://doi.org/10.1103/PhysRevD.104.063523} {\bibfield
  {journal} {\bibinfo  {journal} {Phys. Rev. D}\ }\textbf {\bibinfo {volume}
  {104}},\ \bibinfo {pages} {063523} (\bibinfo {year} {2021})},\ \Eprint
  {https://arxiv.org/abs/2012.11830} {arXiv:2012.11830 [astro-ph.CO]}
  \BibitemShut {NoStop}%
\bibitem [{\citenamefont {Di~Valentino}\ \emph {et~al.}(2025)\citenamefont
  {Di~Valentino} \emph {et~al.}}]{DiValentino:2025sru}%
  \BibitemOpen
  \bibfield  {author} {\bibinfo {author} {\bibfnamefont {E.}~\bibnamefont
  {Di~Valentino}} \emph {et~al.},\ }\href@noop {} {\  (\bibinfo {year}
  {2025})},\ \Eprint {https://arxiv.org/abs/2504.01669} {arXiv:2504.01669
  [astro-ph.CO]} \BibitemShut {NoStop}%
\bibitem [{\citenamefont {Elbers}\ \emph {et~al.}(2025)\citenamefont {Elbers}
  \emph {et~al.}}]{DESI:2025ejh}%
  \BibitemOpen
  \bibfield  {author} {\bibinfo {author} {\bibfnamefont {W.}~\bibnamefont
  {Elbers}} \emph {et~al.} (\bibinfo {collaboration} {DESI}),\ }\href@noop {}
  {\  (\bibinfo {year} {2025})},\ \Eprint {https://arxiv.org/abs/2503.14744}
  {arXiv:2503.14744 [astro-ph.CO]} \BibitemShut {NoStop}%
\bibitem [{\citenamefont {Poudou}\ \emph {et~al.}(2025)\citenamefont {Poudou},
  \citenamefont {Simon}, \citenamefont {Montandon}, \citenamefont {Teixeira},\
  and\ \citenamefont {Poulin}}]{Poudou:2025qcx}%
  \BibitemOpen
  \bibfield  {author} {\bibinfo {author} {\bibfnamefont {A.}~\bibnamefont
  {Poudou}}, \bibinfo {author} {\bibfnamefont {T.}~\bibnamefont {Simon}},
  \bibinfo {author} {\bibfnamefont {T.}~\bibnamefont {Montandon}}, \bibinfo
  {author} {\bibfnamefont {E.~M.}\ \bibnamefont {Teixeira}},\ and\ \bibinfo
  {author} {\bibfnamefont {V.}~\bibnamefont {Poulin}},\ }\href@noop {} {\
  (\bibinfo {year} {2025})},\ \Eprint {https://arxiv.org/abs/2503.10485}
  {arXiv:2503.10485 [astro-ph.CO]} \BibitemShut {NoStop}%
\bibitem [{\citenamefont {Fuller}\ \emph {et~al.}(1988)\citenamefont {Fuller},
  \citenamefont {Mayle},\ and\ \citenamefont {Wilson}}]{Fuller:1988ega}%
  \BibitemOpen
  \bibfield  {author} {\bibinfo {author} {\bibfnamefont {G.~M.}\ \bibnamefont
  {Fuller}}, \bibinfo {author} {\bibfnamefont {R.}~\bibnamefont {Mayle}},\ and\
  \bibinfo {author} {\bibfnamefont {J.~R.}\ \bibnamefont {Wilson}},\ }\href
  {https://doi.org/10.1086/166695} {\bibfield  {journal} {\bibinfo  {journal}
  {Astrophys. J.}\ }\textbf {\bibinfo {volume} {332}},\ \bibinfo {pages} {826}
  (\bibinfo {year} {1988})}\BibitemShut {NoStop}%
\bibitem [{\citenamefont {Kachelriess}\ \emph {et~al.}(2000)\citenamefont
  {Kachelriess}, \citenamefont {Tomas},\ and\ \citenamefont
  {Valle}}]{Kachelriess:2000qc}%
  \BibitemOpen
  \bibfield  {author} {\bibinfo {author} {\bibfnamefont {M.}~\bibnamefont
  {Kachelriess}}, \bibinfo {author} {\bibfnamefont {R.}~\bibnamefont {Tomas}},\
  and\ \bibinfo {author} {\bibfnamefont {J.~W.~F.}\ \bibnamefont {Valle}},\
  }\href {https://doi.org/10.1103/PhysRevD.62.023004} {\bibfield  {journal}
  {\bibinfo  {journal} {Phys. Rev. D}\ }\textbf {\bibinfo {volume} {62}},\
  \bibinfo {pages} {023004} (\bibinfo {year} {2000})},\ \Eprint
  {https://arxiv.org/abs/hep-ph/0001039} {arXiv:hep-ph/0001039} \BibitemShut
  {NoStop}%
\bibitem [{\citenamefont {Farzan}(2003)}]{Farzan:2002wx}%
  \BibitemOpen
  \bibfield  {author} {\bibinfo {author} {\bibfnamefont {Y.}~\bibnamefont
  {Farzan}},\ }\href {https://doi.org/10.1103/PhysRevD.67.073015} {\bibfield
  {journal} {\bibinfo  {journal} {Phys. Rev. D}\ }\textbf {\bibinfo {volume}
  {67}},\ \bibinfo {pages} {073015} (\bibinfo {year} {2003})},\ \Eprint
  {https://arxiv.org/abs/hep-ph/0211375} {arXiv:hep-ph/0211375} \BibitemShut
  {NoStop}%
\bibitem [{\citenamefont {Das}\ \emph {et~al.}(2017)\citenamefont {Das},
  \citenamefont {Dighe},\ and\ \citenamefont {Sen}}]{Das:2017iuj}%
  \BibitemOpen
  \bibfield  {author} {\bibinfo {author} {\bibfnamefont {A.}~\bibnamefont
  {Das}}, \bibinfo {author} {\bibfnamefont {A.}~\bibnamefont {Dighe}},\ and\
  \bibinfo {author} {\bibfnamefont {M.}~\bibnamefont {Sen}},\ }\href
  {https://doi.org/10.1088/1475-7516/2017/05/051} {\bibfield  {journal}
  {\bibinfo  {journal} {JCAP}\ }\textbf {\bibinfo {volume} {05}},\ \bibinfo
  {pages} {051}},\ \Eprint {https://arxiv.org/abs/1705.00468} {arXiv:1705.00468
  [hep-ph]} \BibitemShut {NoStop}%
\bibitem [{\citenamefont {Chang}\ \emph {et~al.}(2023)\citenamefont {Chang},
  \citenamefont {Esteban}, \citenamefont {Beacom}, \citenamefont {Thompson},\
  and\ \citenamefont {Hirata}}]{Chang:2022aas}%
  \BibitemOpen
  \bibfield  {author} {\bibinfo {author} {\bibfnamefont {P.-W.}\ \bibnamefont
  {Chang}}, \bibinfo {author} {\bibfnamefont {I.}~\bibnamefont {Esteban}},
  \bibinfo {author} {\bibfnamefont {J.~F.}\ \bibnamefont {Beacom}}, \bibinfo
  {author} {\bibfnamefont {T.~A.}\ \bibnamefont {Thompson}},\ and\ \bibinfo
  {author} {\bibfnamefont {C.~M.}\ \bibnamefont {Hirata}},\ }\href
  {https://doi.org/10.1103/PhysRevLett.131.071002} {\bibfield  {journal}
  {\bibinfo  {journal} {Phys. Rev. Lett.}\ }\textbf {\bibinfo {volume} {131}},\
  \bibinfo {pages} {071002} (\bibinfo {year} {2023})},\ \Eprint
  {https://arxiv.org/abs/2206.12426} {arXiv:2206.12426 [hep-ph]} \BibitemShut
  {NoStop}%
\bibitem [{\citenamefont {Heurtier}\ and\ \citenamefont
  {Zhang}(2017)}]{Heurtier:2016otg}%
  \BibitemOpen
  \bibfield  {author} {\bibinfo {author} {\bibfnamefont {L.}~\bibnamefont
  {Heurtier}}\ and\ \bibinfo {author} {\bibfnamefont {Y.}~\bibnamefont
  {Zhang}},\ }\href {https://doi.org/10.1088/1475-7516/2017/02/042} {\bibfield
  {journal} {\bibinfo  {journal} {JCAP}\ }\textbf {\bibinfo {volume} {02}},\
  \bibinfo {pages} {042}},\ \Eprint {https://arxiv.org/abs/1609.05882}
  {arXiv:1609.05882 [hep-ph]} \BibitemShut {NoStop}%
\bibitem [{\citenamefont {Fiorillo}\ \emph {et~al.}(2023)\citenamefont
  {Fiorillo}, \citenamefont {Raffelt},\ and\ \citenamefont
  {Vitagliano}}]{Fiorillo:2022cdq}%
  \BibitemOpen
  \bibfield  {author} {\bibinfo {author} {\bibfnamefont {D.~F.~G.}\
  \bibnamefont {Fiorillo}}, \bibinfo {author} {\bibfnamefont {G.~G.}\
  \bibnamefont {Raffelt}},\ and\ \bibinfo {author} {\bibfnamefont
  {E.}~\bibnamefont {Vitagliano}},\ }\href
  {https://doi.org/10.1103/PhysRevLett.131.021001} {\bibfield  {journal}
  {\bibinfo  {journal} {Phys. Rev. Lett.}\ }\textbf {\bibinfo {volume} {131}},\
  \bibinfo {pages} {021001} (\bibinfo {year} {2023})},\ \Eprint
  {https://arxiv.org/abs/2209.11773} {arXiv:2209.11773 [hep-ph]} \BibitemShut
  {NoStop}%
\bibitem [{\citenamefont {Telalovic}\ \emph {et~al.}(2024)\citenamefont
  {Telalovic}, \citenamefont {Fiorillo}, \citenamefont
  {Mart\'\i{}nez-Mirav\'e}, \citenamefont {Vitagliano},\ and\ \citenamefont
  {Bustamante}}]{Telalovic:2024cot}%
  \BibitemOpen
  \bibfield  {author} {\bibinfo {author} {\bibfnamefont {B.}~\bibnamefont
  {Telalovic}}, \bibinfo {author} {\bibfnamefont {D.~F.~G.}\ \bibnamefont
  {Fiorillo}}, \bibinfo {author} {\bibfnamefont {P.}~\bibnamefont
  {Mart\'\i{}nez-Mirav\'e}}, \bibinfo {author} {\bibfnamefont {E.}~\bibnamefont
  {Vitagliano}},\ and\ \bibinfo {author} {\bibfnamefont {M.}~\bibnamefont
  {Bustamante}},\ }\href {https://doi.org/10.1088/1475-7516/2024/11/011}
  {\bibfield  {journal} {\bibinfo  {journal} {JCAP}\ }\textbf {\bibinfo
  {volume} {11}},\ \bibinfo {pages} {011}},\ \Eprint
  {https://arxiv.org/abs/2406.15506} {arXiv:2406.15506 [hep-ph]} \BibitemShut
  {NoStop}%
\bibitem [{\citenamefont {Akita}\ \emph {et~al.}(2022)\citenamefont {Akita},
  \citenamefont {Im},\ and\ \citenamefont {Masud}}]{Akita:2022etk}%
  \BibitemOpen
  \bibfield  {author} {\bibinfo {author} {\bibfnamefont {K.}~\bibnamefont
  {Akita}}, \bibinfo {author} {\bibfnamefont {S.~H.}\ \bibnamefont {Im}},\ and\
  \bibinfo {author} {\bibfnamefont {M.}~\bibnamefont {Masud}},\ }\href
  {https://doi.org/10.1007/JHEP12(2022)050} {\bibfield  {journal} {\bibinfo
  {journal} {JHEP}\ }\textbf {\bibinfo {volume} {12}},\ \bibinfo {pages}
  {050}},\ \Eprint {https://arxiv.org/abs/2206.06852} {arXiv:2206.06852
  [hep-ph]} \BibitemShut {NoStop}%
\bibitem [{\citenamefont {Akita}\ \emph {et~al.}(2024)\citenamefont {Akita},
  \citenamefont {Im}, \citenamefont {Masud},\ and\ \citenamefont
  {Yun}}]{Akita:2023iwq}%
  \BibitemOpen
  \bibfield  {author} {\bibinfo {author} {\bibfnamefont {K.}~\bibnamefont
  {Akita}}, \bibinfo {author} {\bibfnamefont {S.~H.}\ \bibnamefont {Im}},
  \bibinfo {author} {\bibfnamefont {M.}~\bibnamefont {Masud}},\ and\ \bibinfo
  {author} {\bibfnamefont {S.}~\bibnamefont {Yun}},\ }\href
  {https://doi.org/10.1007/JHEP07(2024)057} {\bibfield  {journal} {\bibinfo
  {journal} {JHEP}\ }\textbf {\bibinfo {volume} {07}},\ \bibinfo {pages}
  {057}},\ \Eprint {https://arxiv.org/abs/2312.13627} {arXiv:2312.13627
  [hep-ph]} \BibitemShut {NoStop}%
\bibitem [{\citenamefont {Fiorillo}\ \emph {et~al.}(2024)\citenamefont
  {Fiorillo}, \citenamefont {Raffelt},\ and\ \citenamefont
  {Vitagliano}}]{Fiorillo:2023cas}%
  \BibitemOpen
  \bibfield  {author} {\bibinfo {author} {\bibfnamefont {D.~F.~G.}\
  \bibnamefont {Fiorillo}}, \bibinfo {author} {\bibfnamefont {G.~G.}\
  \bibnamefont {Raffelt}},\ and\ \bibinfo {author} {\bibfnamefont
  {E.}~\bibnamefont {Vitagliano}},\ }\href
  {https://doi.org/10.1103/PhysRevD.109.023017} {\bibfield  {journal} {\bibinfo
   {journal} {Phys. Rev. D}\ }\textbf {\bibinfo {volume} {109}},\ \bibinfo
  {pages} {023017} (\bibinfo {year} {2024})},\ \Eprint
  {https://arxiv.org/abs/2307.15122} {arXiv:2307.15122 [hep-ph]} \BibitemShut
  {NoStop}%
\bibitem [{\citenamefont {Creque-Sarbinowski}\ \emph
  {et~al.}(2021)\citenamefont {Creque-Sarbinowski}, \citenamefont {Hyde},\ and\
  \citenamefont {Kamionkowski}}]{Creque-Sarbinowski:2020qhz}%
  \BibitemOpen
  \bibfield  {author} {\bibinfo {author} {\bibfnamefont {C.}~\bibnamefont
  {Creque-Sarbinowski}}, \bibinfo {author} {\bibfnamefont {J.}~\bibnamefont
  {Hyde}},\ and\ \bibinfo {author} {\bibfnamefont {M.}~\bibnamefont
  {Kamionkowski}},\ }\href {https://doi.org/10.1103/PhysRevD.103.023527}
  {\bibfield  {journal} {\bibinfo  {journal} {Phys. Rev. D}\ }\textbf {\bibinfo
  {volume} {103}},\ \bibinfo {pages} {023527} (\bibinfo {year} {2021})},\
  \Eprint {https://arxiv.org/abs/2005.05332} {arXiv:2005.05332 [hep-ph]}
  \BibitemShut {NoStop}%
\bibitem [{\citenamefont {Shoemaker}\ and\ \citenamefont
  {Murase}(2016)}]{Shoemaker:2015qul}%
  \BibitemOpen
  \bibfield  {author} {\bibinfo {author} {\bibfnamefont {I.~M.}\ \bibnamefont
  {Shoemaker}}\ and\ \bibinfo {author} {\bibfnamefont {K.}~\bibnamefont
  {Murase}},\ }\href {https://doi.org/10.1103/PhysRevD.93.085004} {\bibfield
  {journal} {\bibinfo  {journal} {Phys. Rev. D}\ }\textbf {\bibinfo {volume}
  {93}},\ \bibinfo {pages} {085004} (\bibinfo {year} {2016})},\ \Eprint
  {https://arxiv.org/abs/1512.07228} {arXiv:1512.07228 [astro-ph.HE]}
  \BibitemShut {NoStop}%
\bibitem [{\citenamefont {Fiorillo}\ \emph
  {et~al.}(2020{\natexlab{a}})\citenamefont {Fiorillo}, \citenamefont {Miele},
  \citenamefont {Morisi},\ and\ \citenamefont {Saviano}}]{Fiorillo:2020jvy}%
  \BibitemOpen
  \bibfield  {author} {\bibinfo {author} {\bibfnamefont {D.~F.~G.}\
  \bibnamefont {Fiorillo}}, \bibinfo {author} {\bibfnamefont {G.}~\bibnamefont
  {Miele}}, \bibinfo {author} {\bibfnamefont {S.}~\bibnamefont {Morisi}},\ and\
  \bibinfo {author} {\bibfnamefont {N.}~\bibnamefont {Saviano}},\ }\href
  {https://doi.org/10.1103/PhysRevD.101.083024} {\bibfield  {journal} {\bibinfo
   {journal} {Phys. Rev. D}\ }\textbf {\bibinfo {volume} {101}},\ \bibinfo
  {pages} {083024} (\bibinfo {year} {2020}{\natexlab{a}})},\ \Eprint
  {https://arxiv.org/abs/2002.10125} {arXiv:2002.10125 [hep-ph]} \BibitemShut
  {NoStop}%
\bibitem [{\citenamefont {Fiorillo}\ \emph
  {et~al.}(2020{\natexlab{b}})\citenamefont {Fiorillo}, \citenamefont {Morisi},
  \citenamefont {Miele},\ and\ \citenamefont {Saviano}}]{Fiorillo:2020zzj}%
  \BibitemOpen
  \bibfield  {author} {\bibinfo {author} {\bibfnamefont {D.~F.~G.}\
  \bibnamefont {Fiorillo}}, \bibinfo {author} {\bibfnamefont {S.}~\bibnamefont
  {Morisi}}, \bibinfo {author} {\bibfnamefont {G.}~\bibnamefont {Miele}},\ and\
  \bibinfo {author} {\bibfnamefont {N.}~\bibnamefont {Saviano}},\ }\href
  {https://doi.org/10.1103/PhysRevD.102.083014} {\bibfield  {journal} {\bibinfo
   {journal} {Phys. Rev. D}\ }\textbf {\bibinfo {volume} {102}},\ \bibinfo
  {pages} {083014} (\bibinfo {year} {2020}{\natexlab{b}})},\ \Eprint
  {https://arxiv.org/abs/2007.07866} {arXiv:2007.07866 [hep-ph]} \BibitemShut
  {NoStop}%
\bibitem [{\citenamefont {Ng}\ and\ \citenamefont {Beacom}(2014)}]{Ng:2014pca}%
  \BibitemOpen
  \bibfield  {author} {\bibinfo {author} {\bibfnamefont {K.~C.~Y.}\
  \bibnamefont {Ng}}\ and\ \bibinfo {author} {\bibfnamefont {J.~F.}\
  \bibnamefont {Beacom}},\ }\href {https://doi.org/10.1103/PhysRevD.90.065035}
  {\bibfield  {journal} {\bibinfo  {journal} {Phys. Rev. D}\ }\textbf {\bibinfo
  {volume} {90}},\ \bibinfo {pages} {065035} (\bibinfo {year} {2014})},\
  \bibinfo {note} {[Erratum: Phys.Rev.D 90, 089904 (2014)]},\ \Eprint
  {https://arxiv.org/abs/1404.2288} {arXiv:1404.2288 [astro-ph.HE]}
  \BibitemShut {NoStop}%
\bibitem [{\citenamefont {Farzan}\ and\ \citenamefont
  {Palomares-Ruiz}(2014)}]{Farzan:2014gza}%
  \BibitemOpen
  \bibfield  {author} {\bibinfo {author} {\bibfnamefont {Y.}~\bibnamefont
  {Farzan}}\ and\ \bibinfo {author} {\bibfnamefont {S.}~\bibnamefont
  {Palomares-Ruiz}},\ }\href {https://doi.org/10.1088/1475-7516/2014/06/014}
  {\bibfield  {journal} {\bibinfo  {journal} {JCAP}\ }\textbf {\bibinfo
  {volume} {06}},\ \bibinfo {pages} {014}},\ \Eprint
  {https://arxiv.org/abs/1401.7019} {arXiv:1401.7019 [hep-ph]} \BibitemShut
  {NoStop}%
\bibitem [{\citenamefont {Bustamante}\ \emph {et~al.}(2020)\citenamefont
  {Bustamante}, \citenamefont {Rosenstr\o{}m}, \citenamefont {Shalgar},\ and\
  \citenamefont {Tamborra}}]{Bustamante:2020mep}%
  \BibitemOpen
  \bibfield  {author} {\bibinfo {author} {\bibfnamefont {M.}~\bibnamefont
  {Bustamante}}, \bibinfo {author} {\bibfnamefont {C.}~\bibnamefont
  {Rosenstr\o{}m}}, \bibinfo {author} {\bibfnamefont {S.}~\bibnamefont
  {Shalgar}},\ and\ \bibinfo {author} {\bibfnamefont {I.}~\bibnamefont
  {Tamborra}},\ }\href {https://doi.org/10.1103/PhysRevD.101.123024} {\bibfield
   {journal} {\bibinfo  {journal} {Phys. Rev. D}\ }\textbf {\bibinfo {volume}
  {101}},\ \bibinfo {pages} {123024} (\bibinfo {year} {2020})},\ \Eprint
  {https://arxiv.org/abs/2001.04994} {arXiv:2001.04994 [astro-ph.HE]}
  \BibitemShut {NoStop}%
\bibitem [{\citenamefont {Wang}\ \emph {et~al.}(2025)\citenamefont {Wang},
  \citenamefont {Xu},\ and\ \citenamefont {Zhou}}]{Wang:2025qap}%
  \BibitemOpen
  \bibfield  {author} {\bibinfo {author} {\bibfnamefont {I.~R.}\ \bibnamefont
  {Wang}}, \bibinfo {author} {\bibfnamefont {X.-J.}\ \bibnamefont {Xu}},\ and\
  \bibinfo {author} {\bibfnamefont {B.}~\bibnamefont {Zhou}},\ }\href@noop {}
  {\  (\bibinfo {year} {2025})},\ \Eprint {https://arxiv.org/abs/2501.07624}
  {arXiv:2501.07624 [hep-ph]} \BibitemShut {NoStop}%
\bibitem [{\citenamefont {Mazumdar}\ \emph {et~al.}(2022)\citenamefont
  {Mazumdar}, \citenamefont {Mohanty},\ and\ \citenamefont
  {Parashari}}]{Mazumdar:2020ibx}%
  \BibitemOpen
  \bibfield  {author} {\bibinfo {author} {\bibfnamefont {A.}~\bibnamefont
  {Mazumdar}}, \bibinfo {author} {\bibfnamefont {S.}~\bibnamefont {Mohanty}},\
  and\ \bibinfo {author} {\bibfnamefont {P.}~\bibnamefont {Parashari}},\ }\href
  {https://doi.org/10.1088/1475-7516/2022/10/011} {\bibfield  {journal}
  {\bibinfo  {journal} {JCAP}\ }\textbf {\bibinfo {volume} {10}},\ \bibinfo
  {pages} {011}},\ \Eprint {https://arxiv.org/abs/2011.13685} {arXiv:2011.13685
  [hep-ph]} \BibitemShut {NoStop}%
\bibitem [{\citenamefont {Leal}\ \emph {et~al.}(2025)\citenamefont {Leal},
  \citenamefont {Naredo-Tuero},\ and\ \citenamefont {Funchal}}]{Leal:2025eou}%
  \BibitemOpen
  \bibfield  {author} {\bibinfo {author} {\bibfnamefont {L.~P.~S.}\
  \bibnamefont {Leal}}, \bibinfo {author} {\bibfnamefont {D.}~\bibnamefont
  {Naredo-Tuero}},\ and\ \bibinfo {author} {\bibfnamefont {R.~Z.}\ \bibnamefont
  {Funchal}},\ }\href@noop {} {\  (\bibinfo {year} {2025})},\ \Eprint
  {https://arxiv.org/abs/2504.10576} {arXiv:2504.10576 [hep-ph]} \BibitemShut
  {NoStop}%
\bibitem [{\citenamefont {Borah}\ \emph {et~al.}(2025)\citenamefont {Borah},
  \citenamefont {Das}, \citenamefont {Okada},\ and\ \citenamefont
  {Sarmah}}]{Borah:2025igh}%
  \BibitemOpen
  \bibfield  {author} {\bibinfo {author} {\bibfnamefont {D.}~\bibnamefont
  {Borah}}, \bibinfo {author} {\bibfnamefont {N.}~\bibnamefont {Das}}, \bibinfo
  {author} {\bibfnamefont {N.}~\bibnamefont {Okada}},\ and\ \bibinfo {author}
  {\bibfnamefont {P.}~\bibnamefont {Sarmah}},\ }\href@noop {} {\  (\bibinfo
  {year} {2025})},\ \Eprint {https://arxiv.org/abs/2503.00097}
  {arXiv:2503.00097 [hep-ph]} \BibitemShut {NoStop}%
\bibitem [{\citenamefont {D\"oring}\ and\ \citenamefont
  {Vogl}(2024)}]{Doring:2023vmk}%
  \BibitemOpen
  \bibfield  {author} {\bibinfo {author} {\bibfnamefont {C.}~\bibnamefont
  {D\"oring}}\ and\ \bibinfo {author} {\bibfnamefont {S.}~\bibnamefont
  {Vogl}},\ }\href {https://doi.org/10.1088/1475-7516/2024/07/015} {\bibfield
  {journal} {\bibinfo  {journal} {JCAP}\ }\textbf {\bibinfo {volume} {07}},\
  \bibinfo {pages} {015}},\ \Eprint {https://arxiv.org/abs/2304.08533}
  {arXiv:2304.08533 [hep-ph]} \BibitemShut {NoStop}%
\bibitem [{\citenamefont {Das}\ \emph {et~al.}(2022)\citenamefont {Das},
  \citenamefont {Perez-Gonzalez},\ and\ \citenamefont {Sen}}]{Das:2022xsz}%
  \BibitemOpen
  \bibfield  {author} {\bibinfo {author} {\bibfnamefont {A.}~\bibnamefont
  {Das}}, \bibinfo {author} {\bibfnamefont {Y.~F.}\ \bibnamefont
  {Perez-Gonzalez}},\ and\ \bibinfo {author} {\bibfnamefont {M.}~\bibnamefont
  {Sen}},\ }\href {https://doi.org/10.1103/PhysRevD.106.095042} {\bibfield
  {journal} {\bibinfo  {journal} {Phys. Rev. D}\ }\textbf {\bibinfo {volume}
  {106}},\ \bibinfo {pages} {095042} (\bibinfo {year} {2022})},\ \Eprint
  {https://arxiv.org/abs/2204.11885} {arXiv:2204.11885 [hep-ph]} \BibitemShut
  {NoStop}%
\bibitem [{\citenamefont {Li}\ and\ \citenamefont {Xu}(2023)}]{Li:2023puz}%
  \BibitemOpen
  \bibfield  {author} {\bibinfo {author} {\bibfnamefont {S.-P.}\ \bibnamefont
  {Li}}\ and\ \bibinfo {author} {\bibfnamefont {X.-J.}\ \bibnamefont {Xu}},\
  }\href {https://doi.org/10.1007/JHEP10(2023)012} {\bibfield  {journal}
  {\bibinfo  {journal} {JHEP}\ }\textbf {\bibinfo {volume} {10}},\ \bibinfo
  {pages} {012}},\ \Eprint {https://arxiv.org/abs/2307.13967} {arXiv:2307.13967
  [hep-ph]} \BibitemShut {NoStop}%
\bibitem [{\citenamefont {Huang}\ and\ \citenamefont
  {Rodejohann}(2021)}]{Huang:2021dba}%
  \BibitemOpen
  \bibfield  {author} {\bibinfo {author} {\bibfnamefont {G.-y.}\ \bibnamefont
  {Huang}}\ and\ \bibinfo {author} {\bibfnamefont {W.}~\bibnamefont
  {Rodejohann}},\ }\href {https://doi.org/10.1103/PhysRevD.103.123007}
  {\bibfield  {journal} {\bibinfo  {journal} {Phys. Rev. D}\ }\textbf {\bibinfo
  {volume} {103}},\ \bibinfo {pages} {123007} (\bibinfo {year} {2021})},\
  \Eprint {https://arxiv.org/abs/2102.04280} {arXiv:2102.04280 [hep-ph]}
  \BibitemShut {NoStop}%
\bibitem [{\citenamefont {Ioka}\ and\ \citenamefont
  {Murase}(2014)}]{Ioka:2014kca}%
  \BibitemOpen
  \bibfield  {author} {\bibinfo {author} {\bibfnamefont {K.}~\bibnamefont
  {Ioka}}\ and\ \bibinfo {author} {\bibfnamefont {K.}~\bibnamefont {Murase}},\
  }\href {https://doi.org/10.1093/ptep/ptu090} {\bibfield  {journal} {\bibinfo
  {journal} {PTEP}\ }\textbf {\bibinfo {volume} {2014}},\ \bibinfo {pages}
  {061E01} (\bibinfo {year} {2014})},\ \Eprint
  {https://arxiv.org/abs/1404.2279} {arXiv:1404.2279 [astro-ph.HE]}
  \BibitemShut {NoStop}%
\bibitem [{\citenamefont {Deppisch}\ \emph {et~al.}(2020)\citenamefont
  {Deppisch}, \citenamefont {Graf}, \citenamefont {Rodejohann},\ and\
  \citenamefont {Xu}}]{Deppisch:2020sqh}%
  \BibitemOpen
  \bibfield  {author} {\bibinfo {author} {\bibfnamefont {F.~F.}\ \bibnamefont
  {Deppisch}}, \bibinfo {author} {\bibfnamefont {L.}~\bibnamefont {Graf}},
  \bibinfo {author} {\bibfnamefont {W.}~\bibnamefont {Rodejohann}},\ and\
  \bibinfo {author} {\bibfnamefont {X.-J.}\ \bibnamefont {Xu}},\ }\href
  {https://doi.org/10.1103/PhysRevD.102.051701} {\bibfield  {journal} {\bibinfo
   {journal} {Phys. Rev. D}\ }\textbf {\bibinfo {volume} {102}},\ \bibinfo
  {pages} {051701} (\bibinfo {year} {2020})},\ \Eprint
  {https://arxiv.org/abs/2004.11919} {arXiv:2004.11919 [hep-ph]} \BibitemShut
  {NoStop}%
\bibitem [{\citenamefont {Brdar}\ \emph {et~al.}(2020)\citenamefont {Brdar},
  \citenamefont {Lindner}, \citenamefont {Vogl},\ and\ \citenamefont
  {Xu}}]{Brdar:2020nbj}%
  \BibitemOpen
  \bibfield  {author} {\bibinfo {author} {\bibfnamefont {V.}~\bibnamefont
  {Brdar}}, \bibinfo {author} {\bibfnamefont {M.}~\bibnamefont {Lindner}},
  \bibinfo {author} {\bibfnamefont {S.}~\bibnamefont {Vogl}},\ and\ \bibinfo
  {author} {\bibfnamefont {X.-J.}\ \bibnamefont {Xu}},\ }\href
  {https://doi.org/10.1103/PhysRevD.101.115001} {\bibfield  {journal} {\bibinfo
   {journal} {Phys. Rev. D}\ }\textbf {\bibinfo {volume} {101}},\ \bibinfo
  {pages} {115001} (\bibinfo {year} {2020})},\ \Eprint
  {https://arxiv.org/abs/2003.05339} {arXiv:2003.05339 [hep-ph]} \BibitemShut
  {NoStop}%
\bibitem [{\citenamefont {Dev}\ \emph {et~al.}(2024)\citenamefont {Dev},
  \citenamefont {Kim}, \citenamefont {Sathyan}, \citenamefont {Sinha},\ and\
  \citenamefont {Zhang}}]{Dev:2024twk}%
  \BibitemOpen
  \bibfield  {author} {\bibinfo {author} {\bibfnamefont {P.~S.~B.}\
  \bibnamefont {Dev}}, \bibinfo {author} {\bibfnamefont {D.}~\bibnamefont
  {Kim}}, \bibinfo {author} {\bibfnamefont {D.}~\bibnamefont {Sathyan}},
  \bibinfo {author} {\bibfnamefont {K.}~\bibnamefont {Sinha}},\ and\ \bibinfo
  {author} {\bibfnamefont {Y.}~\bibnamefont {Zhang}},\ }\href@noop {} {\
  (\bibinfo {year} {2024})},\ \Eprint {https://arxiv.org/abs/2407.12738}
  {arXiv:2407.12738 [hep-ph]} \BibitemShut {NoStop}%
\bibitem [{\citenamefont {Aiello}\ \emph {et~al.}(2025)\citenamefont {Aiello}
  \emph {et~al.}}]{KM3NeT:2025npi}%
  \BibitemOpen
  \bibfield  {author} {\bibinfo {author} {\bibfnamefont {S.}~\bibnamefont
  {Aiello}} \emph {et~al.} (\bibinfo {collaboration} {KM3NeT}),\ }\href
  {https://doi.org/10.1038/s41586-024-08543-1} {\bibfield  {journal} {\bibinfo
  {journal} {Nature}\ }\textbf {\bibinfo {volume} {638}},\ \bibinfo {pages}
  {376} (\bibinfo {year} {2025})}\BibitemShut {NoStop}%
\bibitem [{\citenamefont {Jho}\ \emph {et~al.}(2025)\citenamefont {Jho},
  \citenamefont {Park},\ and\ \citenamefont {Shin}}]{Jho:2025gaf}%
  \BibitemOpen
  \bibfield  {author} {\bibinfo {author} {\bibfnamefont {Y.}~\bibnamefont
  {Jho}}, \bibinfo {author} {\bibfnamefont {S.~C.}\ \bibnamefont {Park}},\ and\
  \bibinfo {author} {\bibfnamefont {C.~S.}\ \bibnamefont {Shin}},\ }\href@noop
  {} {\  (\bibinfo {year} {2025})},\ \Eprint {https://arxiv.org/abs/2503.18737}
  {arXiv:2503.18737 [hep-ph]} \BibitemShut {NoStop}%
\bibitem [{\citenamefont {Jiang}\ and\ \citenamefont
  {Huang}(2025)}]{Jiang:2025blz}%
  \BibitemOpen
  \bibfield  {author} {\bibinfo {author} {\bibfnamefont {S.}~\bibnamefont
  {Jiang}}\ and\ \bibinfo {author} {\bibfnamefont {F.~P.}\ \bibnamefont
  {Huang}},\ }\href@noop {} {\  (\bibinfo {year} {2025})},\ \Eprint
  {https://arxiv.org/abs/2503.14332} {arXiv:2503.14332 [hep-ph]} \BibitemShut
  {NoStop}%
\bibitem [{\citenamefont {Kohri}\ \emph {et~al.}(2025)\citenamefont {Kohri},
  \citenamefont {Paul},\ and\ \citenamefont {Sahu}}]{Kohri:2025bsn}%
  \BibitemOpen
  \bibfield  {author} {\bibinfo {author} {\bibfnamefont {K.}~\bibnamefont
  {Kohri}}, \bibinfo {author} {\bibfnamefont {P.~K.}\ \bibnamefont {Paul}},\
  and\ \bibinfo {author} {\bibfnamefont {N.}~\bibnamefont {Sahu}},\ }\href@noop
  {} {\  (\bibinfo {year} {2025})},\ \Eprint {https://arxiv.org/abs/2503.04464}
  {arXiv:2503.04464 [hep-ph]} \BibitemShut {NoStop}%
\bibitem [{\citenamefont {Khan}\ \emph {et~al.}(2025)\citenamefont {Khan},
  \citenamefont {Kim},\ and\ \citenamefont {Ko}}]{Khan:2025gxs}%
  \BibitemOpen
  \bibfield  {author} {\bibinfo {author} {\bibfnamefont {S.}~\bibnamefont
  {Khan}}, \bibinfo {author} {\bibfnamefont {J.}~\bibnamefont {Kim}},\ and\
  \bibinfo {author} {\bibfnamefont {P.}~\bibnamefont {Ko}},\ }\href@noop {} {\
  (\bibinfo {year} {2025})},\ \Eprint {https://arxiv.org/abs/2504.16040}
  {arXiv:2504.16040 [hep-ph]} \BibitemShut {NoStop}%
\bibitem [{\citenamefont {Murase}\ \emph {et~al.}(2025)\citenamefont {Murase},
  \citenamefont {Narita},\ and\ \citenamefont {Yin}}]{Murase:2025uwv}%
  \BibitemOpen
  \bibfield  {author} {\bibinfo {author} {\bibfnamefont {K.}~\bibnamefont
  {Murase}}, \bibinfo {author} {\bibfnamefont {Y.}~\bibnamefont {Narita}},\
  and\ \bibinfo {author} {\bibfnamefont {W.}~\bibnamefont {Yin}},\ }\href@noop
  {} {\  (\bibinfo {year} {2025})},\ \Eprint {https://arxiv.org/abs/2504.15272}
  {arXiv:2504.15272 [hep-ph]} \BibitemShut {NoStop}%
\bibitem [{\citenamefont {Barman}\ \emph {et~al.}(2025)\citenamefont {Barman},
  \citenamefont {Das},\ and\ \citenamefont {Sarmah}}]{Barman:2025hoz}%
  \BibitemOpen
  \bibfield  {author} {\bibinfo {author} {\bibfnamefont {B.}~\bibnamefont
  {Barman}}, \bibinfo {author} {\bibfnamefont {A.}~\bibnamefont {Das}},\ and\
  \bibinfo {author} {\bibfnamefont {P.}~\bibnamefont {Sarmah}},\ }\href@noop {}
  {\  (\bibinfo {year} {2025})},\ \Eprint {https://arxiv.org/abs/2504.01447}
  {arXiv:2504.01447 [hep-ph]} \BibitemShut {NoStop}%
\bibitem [{\citenamefont {Alves}\ \emph {et~al.}(2025)\citenamefont {Alves},
  \citenamefont {Hostert},\ and\ \citenamefont {Pospelov}}]{Alves:2025xul}%
  \BibitemOpen
  \bibfield  {author} {\bibinfo {author} {\bibfnamefont {G.~F.~S.}\
  \bibnamefont {Alves}}, \bibinfo {author} {\bibfnamefont {M.}~\bibnamefont
  {Hostert}},\ and\ \bibinfo {author} {\bibfnamefont {M.}~\bibnamefont
  {Pospelov}},\ }\href@noop {} {\  (\bibinfo {year} {2025})},\ \Eprint
  {https://arxiv.org/abs/2503.14419} {arXiv:2503.14419 [hep-ph]} \BibitemShut
  {NoStop}%
\bibitem [{\citenamefont {Narita}\ and\ \citenamefont
  {Yin}(2025)}]{Narita:2025udw}%
  \BibitemOpen
  \bibfield  {author} {\bibinfo {author} {\bibfnamefont {Y.}~\bibnamefont
  {Narita}}\ and\ \bibinfo {author} {\bibfnamefont {W.}~\bibnamefont {Yin}},\
  }\href@noop {} {\  (\bibinfo {year} {2025})},\ \Eprint
  {https://arxiv.org/abs/2503.07776} {arXiv:2503.07776 [hep-ph]} \BibitemShut
  {NoStop}%
\bibitem [{\citenamefont {Klipfel}\ and\ \citenamefont
  {Kaiser}(2025)}]{Klipfel:2025jql}%
  \BibitemOpen
  \bibfield  {author} {\bibinfo {author} {\bibfnamefont {A.~P.}\ \bibnamefont
  {Klipfel}}\ and\ \bibinfo {author} {\bibfnamefont {D.~I.}\ \bibnamefont
  {Kaiser}},\ }\href@noop {} {\  (\bibinfo {year} {2025})},\ \Eprint
  {https://arxiv.org/abs/2503.19227} {arXiv:2503.19227 [hep-ph]} \BibitemShut
  {NoStop}%
\bibitem [{\citenamefont {Choi}\ \emph {et~al.}(2025)\citenamefont {Choi},
  \citenamefont {Lkhagvadorj},\ and\ \citenamefont {Mahapatra}}]{Choi:2025hqt}%
  \BibitemOpen
  \bibfield  {author} {\bibinfo {author} {\bibfnamefont {K.-Y.}\ \bibnamefont
  {Choi}}, \bibinfo {author} {\bibfnamefont {E.}~\bibnamefont {Lkhagvadorj}},\
  and\ \bibinfo {author} {\bibfnamefont {S.}~\bibnamefont {Mahapatra}},\
  }\href@noop {} {\  (\bibinfo {year} {2025})},\ \Eprint
  {https://arxiv.org/abs/2503.22465} {arXiv:2503.22465 [hep-ph]} \BibitemShut
  {NoStop}%
\bibitem [{\citenamefont {Adriani}\ \emph
  {et~al.}(2025{\natexlab{a}})\citenamefont {Adriani} \emph
  {et~al.}}]{KM3NeT:2025bxl}%
  \BibitemOpen
  \bibfield  {author} {\bibinfo {author} {\bibfnamefont {O.}~\bibnamefont
  {Adriani}} \emph {et~al.} (\bibinfo {collaboration} {KM3NeT, MessMapp Group,
  Fermi-LAT, Owens Valley Radio Observatory 40-m Telescope Group, SVOM}),\
  }\href@noop {} {\  (\bibinfo {year} {2025}{\natexlab{a}})},\ \Eprint
  {https://arxiv.org/abs/2502.08484} {arXiv:2502.08484 [astro-ph.HE]}
  \BibitemShut {NoStop}%
\bibitem [{\citenamefont {Dzhatdoev}(2025)}]{Dzhatdoev:2025sdi}%
  \BibitemOpen
  \bibfield  {author} {\bibinfo {author} {\bibfnamefont {T.~A.}\ \bibnamefont
  {Dzhatdoev}},\ }\href@noop {} {\  (\bibinfo {year} {2025})},\ \Eprint
  {https://arxiv.org/abs/2502.11434} {arXiv:2502.11434 [astro-ph.HE]}
  \BibitemShut {NoStop}%
\bibitem [{\citenamefont {Neronov}\ \emph {et~al.}(2025)\citenamefont
  {Neronov}, \citenamefont {Oikonomou},\ and\ \citenamefont
  {Semikoz}}]{Neronov:2025jfj}%
  \BibitemOpen
  \bibfield  {author} {\bibinfo {author} {\bibfnamefont {A.}~\bibnamefont
  {Neronov}}, \bibinfo {author} {\bibfnamefont {F.}~\bibnamefont {Oikonomou}},\
  and\ \bibinfo {author} {\bibfnamefont {D.}~\bibnamefont {Semikoz}},\
  }\href@noop {} {\  (\bibinfo {year} {2025})},\ \Eprint
  {https://arxiv.org/abs/2502.12986} {arXiv:2502.12986 [astro-ph.HE]}
  \BibitemShut {NoStop}%
\bibitem [{\citenamefont {Zhang}\ \emph {et~al.}(2025)\citenamefont {Zhang},
  \citenamefont {Huang},\ and\ \citenamefont {Li}}]{Zhang:2025abk}%
  \BibitemOpen
  \bibfield  {author} {\bibinfo {author} {\bibfnamefont {Q.}~\bibnamefont
  {Zhang}}, \bibinfo {author} {\bibfnamefont {T.-Q.}\ \bibnamefont {Huang}},\
  and\ \bibinfo {author} {\bibfnamefont {Z.}~\bibnamefont {Li}},\ }\href@noop
  {} {\  (\bibinfo {year} {2025})},\ \Eprint {https://arxiv.org/abs/2504.10378}
  {arXiv:2504.10378 [astro-ph.HE]} \BibitemShut {NoStop}%
\bibitem [{\citenamefont {Brdar}\ and\ \citenamefont
  {Chattopadhyay}(2025)}]{Brdar:2025azm}%
  \BibitemOpen
  \bibfield  {author} {\bibinfo {author} {\bibfnamefont {V.}~\bibnamefont
  {Brdar}}\ and\ \bibinfo {author} {\bibfnamefont {D.~S.}\ \bibnamefont
  {Chattopadhyay}},\ }\href@noop {} {\  (\bibinfo {year} {2025})},\ \Eprint
  {https://arxiv.org/abs/2502.21299} {arXiv:2502.21299 [hep-ph]} \BibitemShut
  {NoStop}%
\bibitem [{\citenamefont {Crnogor\v{c}evi\'c}\ \emph
  {et~al.}(2025)\citenamefont {Crnogor\v{c}evi\'c}, \citenamefont {Blanco},\
  and\ \citenamefont {Linden}}]{Crnogorcevic:2025vou}%
  \BibitemOpen
  \bibfield  {author} {\bibinfo {author} {\bibfnamefont {M.}~\bibnamefont
  {Crnogor\v{c}evi\'c}}, \bibinfo {author} {\bibfnamefont {C.}~\bibnamefont
  {Blanco}},\ and\ \bibinfo {author} {\bibfnamefont {T.}~\bibnamefont
  {Linden}},\ }\href@noop {} {\  (\bibinfo {year} {2025})},\ \Eprint
  {https://arxiv.org/abs/2503.16606} {arXiv:2503.16606 [astro-ph.HE]}
  \BibitemShut {NoStop}%
\bibitem [{\citenamefont {Fang}\ \emph {et~al.}(2025)\citenamefont {Fang},
  \citenamefont {Halzen},\ and\ \citenamefont {Hooper}}]{Fang:2025nzg}%
  \BibitemOpen
  \bibfield  {author} {\bibinfo {author} {\bibfnamefont {K.}~\bibnamefont
  {Fang}}, \bibinfo {author} {\bibfnamefont {F.}~\bibnamefont {Halzen}},\ and\
  \bibinfo {author} {\bibfnamefont {D.}~\bibnamefont {Hooper}},\ }\href
  {https://doi.org/10.3847/2041-8213/adbbec} {\bibfield  {journal} {\bibinfo
  {journal} {Astrophys. J. Lett.}\ }\textbf {\bibinfo {volume} {982}},\
  \bibinfo {pages} {L16} (\bibinfo {year} {2025})},\ \Eprint
  {https://arxiv.org/abs/2502.09545} {arXiv:2502.09545 [astro-ph.HE]}
  \BibitemShut {NoStop}%
\bibitem [{\citenamefont {Adriani}\ \emph
  {et~al.}(2025{\natexlab{b}})\citenamefont {Adriani} \emph
  {et~al.}}]{KM3NeT:2025aps}%
  \BibitemOpen
  \bibfield  {author} {\bibinfo {author} {\bibfnamefont {O.}~\bibnamefont
  {Adriani}} \emph {et~al.} (\bibinfo {collaboration} {KM3NeT}),\ }\href@noop
  {} {\  (\bibinfo {year} {2025}{\natexlab{b}})},\ \Eprint
  {https://arxiv.org/abs/2502.08387} {arXiv:2502.08387 [astro-ph.HE]}
  \BibitemShut {NoStop}%
\bibitem [{\citenamefont {Das}\ \emph {et~al.}(2025)\citenamefont {Das},
  \citenamefont {Zhang}, \citenamefont {Razzaque},\ and\ \citenamefont
  {Xu}}]{Das:2025vqd}%
  \BibitemOpen
  \bibfield  {author} {\bibinfo {author} {\bibfnamefont {S.}~\bibnamefont
  {Das}}, \bibinfo {author} {\bibfnamefont {B.}~\bibnamefont {Zhang}}, \bibinfo
  {author} {\bibfnamefont {S.}~\bibnamefont {Razzaque}},\ and\ \bibinfo
  {author} {\bibfnamefont {S.}~\bibnamefont {Xu}},\ }\href@noop {} {\
  (\bibinfo {year} {2025})},\ \Eprint {https://arxiv.org/abs/2504.10847}
  {arXiv:2504.10847 [astro-ph.HE]} \BibitemShut {NoStop}%
\bibitem [{\citenamefont {Yang}\ \emph {et~al.}(2025)\citenamefont {Yang},
  \citenamefont {Lv}, \citenamefont {Bi},\ and\ \citenamefont
  {Yin}}]{Yang:2025kfr}%
  \BibitemOpen
  \bibfield  {author} {\bibinfo {author} {\bibfnamefont {Y.-M.}\ \bibnamefont
  {Yang}}, \bibinfo {author} {\bibfnamefont {X.-J.}\ \bibnamefont {Lv}},
  \bibinfo {author} {\bibfnamefont {X.-J.}\ \bibnamefont {Bi}},\ and\ \bibinfo
  {author} {\bibfnamefont {P.-F.}\ \bibnamefont {Yin}},\ }\href@noop {} {\
  (\bibinfo {year} {2025})},\ \Eprint {https://arxiv.org/abs/2502.18256}
  {arXiv:2502.18256 [hep-ph]} \BibitemShut {NoStop}%
\bibitem [{\citenamefont {Cattaneo}(2025)}]{Cattaneo:2025uxk}%
  \BibitemOpen
  \bibfield  {author} {\bibinfo {author} {\bibfnamefont {P.~W.}\ \bibnamefont
  {Cattaneo}},\ }\href@noop {} {\  (\bibinfo {year} {2025})},\ \Eprint
  {https://arxiv.org/abs/2502.13201} {arXiv:2502.13201 [hep-ph]} \BibitemShut
  {NoStop}%
\bibitem [{\citenamefont {Adriani}\ \emph
  {et~al.}(2025{\natexlab{c}})\citenamefont {Adriani} \emph
  {et~al.}}]{KM3NeT:2025mfl}%
  \BibitemOpen
  \bibfield  {author} {\bibinfo {author} {\bibfnamefont {O.}~\bibnamefont
  {Adriani}} \emph {et~al.} (\bibinfo {collaboration} {KM3NeT}),\ }\href@noop
  {} {\  (\bibinfo {year} {2025}{\natexlab{c}})},\ \Eprint
  {https://arxiv.org/abs/2502.12070} {arXiv:2502.12070 [astro-ph.HE]}
  \BibitemShut {NoStop}%
\bibitem [{\citenamefont {Abbasi}\ \emph {et~al.}(2024)\citenamefont {Abbasi}
  \emph {et~al.}}]{IceCube:2024fxo}%
  \BibitemOpen
  \bibfield  {author} {\bibinfo {author} {\bibfnamefont {R.}~\bibnamefont
  {Abbasi}} \emph {et~al.} (\bibinfo {collaboration} {IceCube}),\ }\href
  {https://doi.org/10.1103/PhysRevD.110.022001} {\bibfield  {journal} {\bibinfo
   {journal} {Phys. Rev. D}\ }\textbf {\bibinfo {volume} {110}},\ \bibinfo
  {pages} {022001} (\bibinfo {year} {2024})},\ \Eprint
  {https://arxiv.org/abs/2402.18026} {arXiv:2402.18026 [astro-ph.HE]}
  \BibitemShut {NoStop}%
\bibitem [{\citenamefont {Abdul~Halim}\ \emph {et~al.}(2023)\citenamefont
  {Abdul~Halim} \emph {et~al.}}]{PierreAuger:2023pjg}%
  \BibitemOpen
  \bibfield  {author} {\bibinfo {author} {\bibfnamefont {A.}~\bibnamefont
  {Abdul~Halim}} \emph {et~al.} (\bibinfo {collaboration} {Pierre Auger}),\
  }\href {https://doi.org/10.22323/1.444.1488} {\bibfield  {journal} {\bibinfo
  {journal} {PoS}\ }\textbf {\bibinfo {volume} {ICRC2023}},\ \bibinfo {pages}
  {1488} (\bibinfo {year} {2023})}\BibitemShut {NoStop}%
\bibitem [{\citenamefont {Abbasi}\ \emph {et~al.}(2022)\citenamefont {Abbasi}
  \emph {et~al.}}]{Abbasi:2021qfz}%
  \BibitemOpen
  \bibfield  {author} {\bibinfo {author} {\bibfnamefont {R.}~\bibnamefont
  {Abbasi}} \emph {et~al.},\ }\href {https://doi.org/10.3847/1538-4357/ac4d29}
  {\bibfield  {journal} {\bibinfo  {journal} {Astrophys. J.}\ }\textbf
  {\bibinfo {volume} {928}},\ \bibinfo {pages} {50} (\bibinfo {year} {2022})},\
  \Eprint {https://arxiv.org/abs/2111.10299} {arXiv:2111.10299 [astro-ph.HE]}
  \BibitemShut {NoStop}%
\bibitem [{\citenamefont {Abbasi}\ \emph {et~al.}(2021)\citenamefont {Abbasi}
  \emph {et~al.}}]{IceCube:2020wum}%
  \BibitemOpen
  \bibfield  {author} {\bibinfo {author} {\bibfnamefont {R.}~\bibnamefont
  {Abbasi}} \emph {et~al.} (\bibinfo {collaboration} {IceCube}),\ }\href
  {https://doi.org/10.1103/PhysRevD.104.022002} {\bibfield  {journal} {\bibinfo
   {journal} {Phys. Rev. D}\ }\textbf {\bibinfo {volume} {104}},\ \bibinfo
  {pages} {022002} (\bibinfo {year} {2021})},\ \Eprint
  {https://arxiv.org/abs/2011.03545} {arXiv:2011.03545 [astro-ph.HE]}
  \BibitemShut {NoStop}%
\bibitem [{\citenamefont {Abbasi}\ \emph {et~al.}(2025)\citenamefont {Abbasi}
  \emph {et~al.}}]{IceCube:2025ary}%
  \BibitemOpen
  \bibfield  {author} {\bibinfo {author} {\bibfnamefont {R.}~\bibnamefont
  {Abbasi}} \emph {et~al.} (\bibinfo {collaboration} {IceCube}),\ }\href@noop
  {} {\  (\bibinfo {year} {2025})},\ \Eprint {https://arxiv.org/abs/2502.19776}
  {arXiv:2502.19776 [astro-ph.HE]} \BibitemShut {NoStop}%
\bibitem [{\citenamefont {Li}\ \emph {et~al.}(2025)\citenamefont {Li},
  \citenamefont {Machado}, \citenamefont {Naredo-Tuero},\ and\ \citenamefont
  {Schwemberger}}]{Li:2025tqf}%
  \BibitemOpen
  \bibfield  {author} {\bibinfo {author} {\bibfnamefont {S.~W.}\ \bibnamefont
  {Li}}, \bibinfo {author} {\bibfnamefont {P.}~\bibnamefont {Machado}},
  \bibinfo {author} {\bibfnamefont {D.}~\bibnamefont {Naredo-Tuero}},\ and\
  \bibinfo {author} {\bibfnamefont {T.}~\bibnamefont {Schwemberger}},\
  }\href@noop {} {\  (\bibinfo {year} {2025})},\ \Eprint
  {https://arxiv.org/abs/2502.04508} {arXiv:2502.04508 [astro-ph.HE]}
  \BibitemShut {NoStop}%
\bibitem [{\citenamefont {Adriani}\ \emph
  {et~al.}(2025{\natexlab{d}})\citenamefont {Adriani} \emph
  {et~al.}}]{KM3NeT:2025ccp}%
  \BibitemOpen
  \bibfield  {author} {\bibinfo {author} {\bibfnamefont {O.}~\bibnamefont
  {Adriani}} \emph {et~al.} (\bibinfo {collaboration} {KM3NeT}),\ }\href@noop
  {} {\  (\bibinfo {year} {2025}{\natexlab{d}})},\ \Eprint
  {https://arxiv.org/abs/2502.08173} {arXiv:2502.08173 [astro-ph.HE]}
  \BibitemShut {NoStop}%
\bibitem [{\citenamefont {Aartsen}\ \emph {et~al.}(2021)\citenamefont {Aartsen}
  \emph {et~al.}}]{IceCube-Gen2:2020qha}%
  \BibitemOpen
  \bibfield  {author} {\bibinfo {author} {\bibfnamefont {M.~G.}\ \bibnamefont
  {Aartsen}} \emph {et~al.} (\bibinfo {collaboration} {IceCube-Gen2}),\ }\href
  {https://doi.org/10.1088/1361-6471/abbd48} {\bibfield  {journal} {\bibinfo
  {journal} {J. Phys. G}\ }\textbf {\bibinfo {volume} {48}},\ \bibinfo {pages}
  {060501} (\bibinfo {year} {2021})},\ \Eprint
  {https://arxiv.org/abs/2008.04323} {arXiv:2008.04323 [astro-ph.HE]}
  \BibitemShut {NoStop}%
\bibitem [{\citenamefont {Meier}(2024)}]{Meier:2024flg}%
  \BibitemOpen
  \bibfield  {author} {\bibinfo {author} {\bibfnamefont {M.}~\bibnamefont
  {Meier}} (\bibinfo {collaboration} {IceCube})\ }(\bibinfo {year} {2024})\
  \Eprint {https://arxiv.org/abs/2409.01740} {arXiv:2409.01740 [astro-ph.HE]}
  \BibitemShut {NoStop}%
\bibitem [{\citenamefont {Blinov}\ \emph {et~al.}(2019)\citenamefont {Blinov},
  \citenamefont {Kelly}, \citenamefont {Krnjaic},\ and\ \citenamefont
  {McDermott}}]{Blinov:2019gcj}%
  \BibitemOpen
  \bibfield  {author} {\bibinfo {author} {\bibfnamefont {N.}~\bibnamefont
  {Blinov}}, \bibinfo {author} {\bibfnamefont {K.~J.}\ \bibnamefont {Kelly}},
  \bibinfo {author} {\bibfnamefont {G.~Z.}\ \bibnamefont {Krnjaic}},\ and\
  \bibinfo {author} {\bibfnamefont {S.~D.}\ \bibnamefont {McDermott}},\ }\href
  {https://doi.org/10.1103/PhysRevLett.123.191102} {\bibfield  {journal}
  {\bibinfo  {journal} {Phys. Rev. Lett.}\ }\textbf {\bibinfo {volume} {123}},\
  \bibinfo {pages} {191102} (\bibinfo {year} {2019})},\ \Eprint
  {https://arxiv.org/abs/1905.02727} {arXiv:1905.02727 [astro-ph.CO]}
  \BibitemShut {NoStop}%
\bibitem [{\citenamefont {Blum}\ \emph {et~al.}(2014)\citenamefont {Blum},
  \citenamefont {Hook},\ and\ \citenamefont {Murase}}]{Blum:2014ewa}%
  \BibitemOpen
  \bibfield  {author} {\bibinfo {author} {\bibfnamefont {K.}~\bibnamefont
  {Blum}}, \bibinfo {author} {\bibfnamefont {A.}~\bibnamefont {Hook}},\ and\
  \bibinfo {author} {\bibfnamefont {K.}~\bibnamefont {Murase}},\ }\href@noop {}
  {\  (\bibinfo {year} {2014})},\ \Eprint {https://arxiv.org/abs/1408.3799}
  {arXiv:1408.3799 [hep-ph]} \BibitemShut {NoStop}%
\bibitem [{\citenamefont {Berryman}\ \emph {et~al.}(2018)\citenamefont
  {Berryman}, \citenamefont {De~Gouv\^ea}, \citenamefont {Kelly},\ and\
  \citenamefont {Zhang}}]{Berryman:2018ogk}%
  \BibitemOpen
  \bibfield  {author} {\bibinfo {author} {\bibfnamefont {J.~M.}\ \bibnamefont
  {Berryman}}, \bibinfo {author} {\bibfnamefont {A.}~\bibnamefont
  {De~Gouv\^ea}}, \bibinfo {author} {\bibfnamefont {K.~J.}\ \bibnamefont
  {Kelly}},\ and\ \bibinfo {author} {\bibfnamefont {Y.}~\bibnamefont {Zhang}},\
  }\href {https://doi.org/10.1103/PhysRevD.97.075030} {\bibfield  {journal}
  {\bibinfo  {journal} {Phys. Rev. D}\ }\textbf {\bibinfo {volume} {97}},\
  \bibinfo {pages} {075030} (\bibinfo {year} {2018})},\ \Eprint
  {https://arxiv.org/abs/1802.00009} {arXiv:1802.00009 [hep-ph]} \BibitemShut
  {NoStop}%
\bibitem [{\citenamefont {Kelly}\ \emph {et~al.}(2020)\citenamefont {Kelly},
  \citenamefont {Sen}, \citenamefont {Tangarife},\ and\ \citenamefont
  {Zhang}}]{Kelly:2020pcy}%
  \BibitemOpen
  \bibfield  {author} {\bibinfo {author} {\bibfnamefont {K.~J.}\ \bibnamefont
  {Kelly}}, \bibinfo {author} {\bibfnamefont {M.}~\bibnamefont {Sen}}, \bibinfo
  {author} {\bibfnamefont {W.}~\bibnamefont {Tangarife}},\ and\ \bibinfo
  {author} {\bibfnamefont {Y.}~\bibnamefont {Zhang}},\ }\href
  {https://doi.org/10.1103/PhysRevD.101.115031} {\bibfield  {journal} {\bibinfo
   {journal} {Phys. Rev. D}\ }\textbf {\bibinfo {volume} {101}},\ \bibinfo
  {pages} {115031} (\bibinfo {year} {2020})},\ \Eprint
  {https://arxiv.org/abs/2005.03681} {arXiv:2005.03681 [hep-ph]} \BibitemShut
  {NoStop}%
\bibitem [{\citenamefont {Esteban}\ \emph {et~al.}(2021)\citenamefont
  {Esteban}, \citenamefont {Pandey}, \citenamefont {Brdar},\ and\ \citenamefont
  {Beacom}}]{Esteban:2021tub}%
  \BibitemOpen
  \bibfield  {author} {\bibinfo {author} {\bibfnamefont {I.}~\bibnamefont
  {Esteban}}, \bibinfo {author} {\bibfnamefont {S.}~\bibnamefont {Pandey}},
  \bibinfo {author} {\bibfnamefont {V.}~\bibnamefont {Brdar}},\ and\ \bibinfo
  {author} {\bibfnamefont {J.~F.}\ \bibnamefont {Beacom}},\ }\href
  {https://doi.org/10.1103/PhysRevD.104.123014} {\bibfield  {journal} {\bibinfo
   {journal} {Phys. Rev. D}\ }\textbf {\bibinfo {volume} {104}},\ \bibinfo
  {pages} {123014} (\bibinfo {year} {2021})},\ \Eprint
  {https://arxiv.org/abs/2107.13568} {arXiv:2107.13568 [hep-ph]} \BibitemShut
  {NoStop}%
\bibitem [{\citenamefont {Aartsen}\ \emph {et~al.}(2018)\citenamefont {Aartsen}
  \emph {et~al.}}]{IceCube:2018fhm}%
  \BibitemOpen
  \bibfield  {author} {\bibinfo {author} {\bibfnamefont {M.~G.}\ \bibnamefont
  {Aartsen}} \emph {et~al.} (\bibinfo {collaboration} {IceCube}),\ }\href
  {https://doi.org/10.1103/PhysRevD.98.062003} {\bibfield  {journal} {\bibinfo
  {journal} {Phys. Rev. D}\ }\textbf {\bibinfo {volume} {98}},\ \bibinfo
  {pages} {062003} (\bibinfo {year} {2018})},\ \Eprint
  {https://arxiv.org/abs/1807.01820} {arXiv:1807.01820 [astro-ph.HE]}
  \BibitemShut {NoStop}%
\bibitem [{Note1()}]{Note1}%
  \BibitemOpen
  \bibinfo {note} {Recent DESI DR2~\cite {DESI:2025ejh} reported a constraint
  on sum of neutrino mass to be $\DOTSB \sum@ \slimits@ m_\nu < 0.0642 \
  \protect \text {eV}$ when assuming $\Lambda $CDM model. But its best fit is
  when allowing dark energy to vary. Under the latter scenario, the neutrino
  total mass constraints relaxed to $\DOTSB \sum@ \slimits@ m_\nu < 0.163 \
  \protect \text {eV}$. This allows us to use 0.1 eV as our benchmark
  value.}\BibitemShut {Stop}%
\bibitem [{\citenamefont {Esteban}\ \emph {et~al.}(2020)\citenamefont
  {Esteban}, \citenamefont {Gonzalez-Garcia}, \citenamefont {Maltoni},
  \citenamefont {Schwetz},\ and\ \citenamefont {Zhou}}]{Esteban:2020cvm}%
  \BibitemOpen
  \bibfield  {author} {\bibinfo {author} {\bibfnamefont {I.}~\bibnamefont
  {Esteban}}, \bibinfo {author} {\bibfnamefont {M.~C.}\ \bibnamefont
  {Gonzalez-Garcia}}, \bibinfo {author} {\bibfnamefont {M.}~\bibnamefont
  {Maltoni}}, \bibinfo {author} {\bibfnamefont {T.}~\bibnamefont {Schwetz}},\
  and\ \bibinfo {author} {\bibfnamefont {A.}~\bibnamefont {Zhou}},\ }\href
  {https://doi.org/10.1007/JHEP09(2020)178} {\bibfield  {journal} {\bibinfo
  {journal} {JHEP}\ }\textbf {\bibinfo {volume} {09}},\ \bibinfo {pages}
  {178}},\ \Eprint {https://arxiv.org/abs/2007.14792} {arXiv:2007.14792
  [hep-ph]} \BibitemShut {NoStop}%
\bibitem [{\citenamefont {Buchner}(2021)}]{Buchner:2021cql}%
  \BibitemOpen
  \bibfield  {author} {\bibinfo {author} {\bibfnamefont {J.}~\bibnamefont
  {Buchner}},\ }\href {https://doi.org/10.21105/joss.03001} {\bibfield
  {journal} {\bibinfo  {journal} {The Journal of Open Source Software}\
  }\textbf {\bibinfo {volume} {6}},\ \bibinfo {pages} {3001} (\bibinfo {year}
  {2021})}\BibitemShut {NoStop}%
\bibitem [{\citenamefont {Lewis}(2019)}]{Lewis:2019xzd}%
  \BibitemOpen
  \bibfield  {author} {\bibinfo {author} {\bibfnamefont {A.}~\bibnamefont
  {Lewis}},\ }\href@noop {} {\  (\bibinfo {year} {2019})},\ \Eprint
  {https://arxiv.org/abs/1910.13970} {arXiv:1910.13970 [astro-ph.IM]}
  \BibitemShut {NoStop}%
\bibitem [{\citenamefont {Albert}\ \emph {et~al.}(2024)\citenamefont {Albert}
  \emph {et~al.}}]{ANTARES:2024ihw}%
  \BibitemOpen
  \bibfield  {author} {\bibinfo {author} {\bibfnamefont {A.}~\bibnamefont
  {Albert}} \emph {et~al.} (\bibinfo {collaboration} {ANTARES}),\ }\href
  {https://doi.org/10.1088/1475-7516/2024/08/038} {\bibfield  {journal}
  {\bibinfo  {journal} {JCAP}\ }\textbf {\bibinfo {volume} {08}},\ \bibinfo
  {pages} {038}},\ \Eprint {https://arxiv.org/abs/2407.00328} {arXiv:2407.00328
  [astro-ph.HE]} \BibitemShut {NoStop}%
\bibitem [{\citenamefont {Kreisch}\ \emph {et~al.}(2020)\citenamefont
  {Kreisch}, \citenamefont {Cyr-Racine},\ and\ \citenamefont
  {Dor\'e}}]{Kreisch:2019yzn}%
  \BibitemOpen
  \bibfield  {author} {\bibinfo {author} {\bibfnamefont {C.~D.}\ \bibnamefont
  {Kreisch}}, \bibinfo {author} {\bibfnamefont {F.-Y.}\ \bibnamefont
  {Cyr-Racine}},\ and\ \bibinfo {author} {\bibfnamefont {O.}~\bibnamefont
  {Dor\'e}},\ }\href {https://doi.org/10.1103/PhysRevD.101.123505} {\bibfield
  {journal} {\bibinfo  {journal} {Phys. Rev. D}\ }\textbf {\bibinfo {volume}
  {101}},\ \bibinfo {pages} {123505} (\bibinfo {year} {2020})},\ \Eprint
  {https://arxiv.org/abs/1902.00534} {arXiv:1902.00534 [astro-ph.CO]}
  \BibitemShut {NoStop}%
\bibitem [{\citenamefont {Camarena}\ and\ \citenamefont
  {Cyr-Racine}(2025)}]{Camarena:2024daj}%
  \BibitemOpen
  \bibfield  {author} {\bibinfo {author} {\bibfnamefont {D.}~\bibnamefont
  {Camarena}}\ and\ \bibinfo {author} {\bibfnamefont {F.-Y.}\ \bibnamefont
  {Cyr-Racine}},\ }\href {https://doi.org/10.1103/PhysRevD.111.023504}
  {\bibfield  {journal} {\bibinfo  {journal} {Phys. Rev. D}\ }\textbf {\bibinfo
  {volume} {111}},\ \bibinfo {pages} {023504} (\bibinfo {year} {2025})},\
  \Eprint {https://arxiv.org/abs/2403.05496} {arXiv:2403.05496 [astro-ph.CO]}
  \BibitemShut {NoStop}%
\bibitem [{\citenamefont {Roy~Choudhury}\ \emph {et~al.}(2021)\citenamefont
  {Roy~Choudhury}, \citenamefont {Hannestad},\ and\ \citenamefont
  {Tram}}]{RoyChoudhury:2020dmd}%
  \BibitemOpen
  \bibfield  {author} {\bibinfo {author} {\bibfnamefont {S.}~\bibnamefont
  {Roy~Choudhury}}, \bibinfo {author} {\bibfnamefont {S.}~\bibnamefont
  {Hannestad}},\ and\ \bibinfo {author} {\bibfnamefont {T.}~\bibnamefont
  {Tram}},\ }\href {https://doi.org/10.1088/1475-7516/2021/03/084} {\bibfield
  {journal} {\bibinfo  {journal} {JCAP}\ }\textbf {\bibinfo {volume} {03}},\
  \bibinfo {pages} {084}},\ \Eprint {https://arxiv.org/abs/2012.07519}
  {arXiv:2012.07519 [astro-ph.CO]} \BibitemShut {NoStop}%
\bibitem [{\citenamefont {Barlow}(2004)}]{Barlow:2004wg}%
  \BibitemOpen
  \bibfield  {author} {\bibinfo {author} {\bibfnamefont {R.}~\bibnamefont
  {Barlow}}\ }(\bibinfo {year} {2004})\ pp.\ \bibinfo {pages} {56--59},\
  \Eprint {https://arxiv.org/abs/physics/0406120} {arXiv:physics/0406120}
  \BibitemShut {NoStop}%
\end{thebibliography}%
\bibliographystyle{apsrev4-2}

\end{document}